\documentclass[numberedappendix,12pt,preprint]{emulateapj}

\usepackage[]{amsmath}

\usepackage{color}			      
\definecolor{midgray}{gray}{0.4}		
\definecolor{orange}{rgb}{1,0.5,0}    

\usepackage[colorlinks=true,citecolor=midgray,linkcolor=midgray]{hyperref}        

\newcommand{\simgt}{\,\rlap{\lower 3.5 pt \hbox{$\mathchar \sim$}} \raise
1pt \hbox {$>$}\,}
\newcommand{\simlt}{\,\rlap{\lower 3.5 pt \hbox{$\mathchar \sim$}} \raise
1pt \hbox {$<$}\,}

\newcommand{\lya}{Ly$\alpha$}

 \newcommand{\BE}{\begin{equation}}
 \newcommand{\EE}{\end{equation}}

\newcommand{\Atwo}{Abell 2744}
\newcommand{\Mseven}{MACS J0717.5+3745}
\newcommand{\Mfourteen}{MACS J1423.8+2404}
\newcommand{\Mtwentyone}{MACS J2129.4-0741}
\newcommand{\Rthirteen}{RXC J1347.5-1145}
\newcommand{\Rtwentytwo}{RXC J2248.7-4431}

\shorttitle{GLASS spectroscopy at $z\gtrsim7$}
\shortauthors{Schmidt et al. (2015)}

\begin{document}

\title{
The Grism Lens-Amplified Survey from Space (GLASS). III. A census of \lya\ Emission at \lowercase{$z\gtrsim7$} from \emph{HST} Spectroscopy}

\author{
K.~B.~Schmidt$^{1,2}$,  
T.~Treu$^{3}$, 
M.~Brada\v{c}$^{4}$, 
B.~Vulcani$^{5}$,
K.-H., Huang$^{4}$,
A.~Hoag$^{4}$, 
M.~Maseda$^{6}$
L.~Guaita$^{7}$
L.~Pentericci$^{7}$, 
G.~B.~Brammer$^{8}$,
M.~Dijkstra$^{9}$,
A.~Dressler$^{10}$
A.~Fontana$^{7}$,
A.~L.~Henry$^{11}$, 
T.~A.~Jones$^{1}$,
C.~Mason$^{1}$,
M.~Trenti$^{12}$, 
X.~Wang$^{1}$,
}
\affil{$^{1}$ Department of Physics, University of California, Santa Barbara, CA, 93106-9530, USA}
\affil{$^{2}$ Leibniz-Institut f\"ur Astrophysik Potsdam (AIP), An der Sternwarte 16, 14482 Potsdam, Germany}
\affil{$^{3}$ Department of Physics and Astronomy, UCLA, Los Angeles, CA, 90095-1547, USA}
\affil{$^{4}$ Department of Physics, University of California, Davis, CA, 95616, USA}
\affil{$^{5}$ Kavli Institute for the Physics and Mathematics of the Universe (WPI), Todai Institutes for Advanced Study, the University of Tokyo, Kashiwa, 277-8582, Japan}
\affil{$^{6}$ Max-Planck-Institut f\"ur Astronomie, K\"onigstuhl 17, D-69117 Heidelberg, Germany}
\affil{$^{7}$ INAF - Osservatorio Astronomico di Roma Via Frascati 33 - 00040 Monte Porzio Catone, I}
\affil{$^{8}$ Space Telescope Science Institute, 3700 San Martin Drive, Baltimore, MD, 21218, USA}
\affil{$^{9}$ Institute of Theoretical Astrophysics, University of Oslo, Postboks 1029, 0858 Oslo, Norway}
\affil{$^{10}$ The Observatories of the Carnegie Institution for Science, 813 Santa Barbara St., Pasadena, CA 91101, USA}
\affil{$^{11}$ Astrophysics Science Division, Goddard Space Flight Center, Code 665, Greenbelt, MD 20771}
\affil{$^{12}$ School of Physics, The University of Melbourne, VIC, 3010 Australia}

\email{kbschmidt@aip.de}

\begin{abstract}
We present a census of \lya\ emission at $z\gtrsim7$ utilizing
deep near infrared \emph{HST} grism spectroscopy from the first six
completed clusters of the Grism Lens-Amplified Survey from Space
(GLASS).  In 24/159 photometrically selected galaxies we detect
emission lines consistent with \lya\ in the GLASS spectra. Based on
the distribution of signal-to-noise ratios and on simulations we
expect the completeness and the purity of the sample to be 40-100\% and
60-90\%, respectively.  For the objects without detected emission lines we show that the
observed (not corrected for lensing magnification) 1$\sigma$
flux limits reaches $5\times10^{-18}$erg/s/cm$^{2}$ per position angle over the full wavelength
range of GLASS (0.8--1.7$\mu$m). Based on the conditional probability
of \lya\ emission measured from the ground at $z\sim7$ we would have
expected 12-18 \lya\ emitters. This is consistent with the number of
detections, within the uncertainties, confirming the drop in \lya\
emission with respect to $z\sim6$.  Deeper follow-up spectroscopy,
here exemplified by Keck spectroscopy, is necessary to improve our
estimates of completeness and purity, and to
confirm individual candidates as true \lya\ emitters.
These candidates include a promising source at $z=8.1$.
The spatial extent of \lya\ in a deep stack of the most convincing
\lya\ emitters with $\langle z\rangle=7.2$ is consistent with that of
the rest-frame UV continuum. Extended \lya\ emission, if present, has
a surface brightness below our detection limit, consistent with the
properties of lower redshift comparison samples.  From the stack we
estimate upper limits on rest-frame UV emission line ratios and find
$f_\textrm{CIV} / f_\textrm{\lya} \lesssim 0.32$ and $f_\textrm{CIII]}
/ f_\textrm{\lya} \lesssim 0.23$ in good agreement with other values
published in the literature.
\end{abstract}


\section{Introduction}
\label{sec:intro}

With the deployment of the wide field camera 3 (WFC3) on the Hubble
Space Telescope (\emph{HST}) in 2009, the samples of galaxies at the
epoch of reionization, the phase-transition from a completely neutral
inter-galactic medium (IGM) to a fully ionized IGM at $z\gtrsim6$,
have grown dramatically.
One of the main results of the WFC3 imaging campaigns has been the
accurate determination of the luminosity function of star forming
high-redshift (based on their photometry) Lyman break galaxies
\citep[e.g.][]{Bouwens:2015p34683,Finkelstein:2015p37430}.
The UV luminosity functions of Lyman break galaxies have provided key constraints on the
physics of reionization
\citep[e.g.][]{Robertson:2013p27340,Schmidt:2014p34189,Duffy:2014p35578}. For example, it is clear that the population of galaxies that has been detected so far cannot produce enough hard photons to keep the universe ionized. However, the luminosity function is found to have a steep faint end slope (approximately $\phi \propto L^{-2}$). 
Thus, faint galaxies could in principle provide enough ionizing photons
\citep{Bouwens:2015p39292,Robertson:2015p38368,BaroneNugent:2015p38474,Dressler:2015p37838}. 
even though a contribution from AGN might end up being necessary \citep{Madau:2015p41145,Giallongo:2015p41144}.

Also ground based spectroscopic follow-up of photometrically selected 
high-redshift candidates has been an important part of these studies and has provided additional clues about the reionization epoch.
Remarkably, only a handful of sources have been confirmed above
redshift 7
\citep{Vanzella:2011p29486,Schenker:2012p34406,Schenker:2014p35145,Ono:2012p27651,Finkelstein:2013p32467,Oesch:2015p38504,RobertsBorsani:2015p39891,Zitrin:2015p40322}. 
The low probability of detecting \lya\ in Lyman break galaxies, could be
interpreted as the result of an increased optical depth in the IGM due to a
significant fraction of neutral hydrogen. Thus the decline in detected \lya\
 is potentially a ``smoking gun'' of reionization
\citep{Fontana:2010p29506}. 
The conditional probability of
\lya\ emission for Lyman break galaxies is potentially a powerful probe of the physics 
of the intergalactic and circumgalactic media an their neutral fraction at these redshifts 
\citep{Dijkstra:2011p33040,Jensen:2013p40647,Dijkstra:2014p39158,Mesinger:2015p39033},
provided that large enough spectroscopic samples can be gathered
\citep{Treu:2012p12658,Treu:2013p32132,Pentericci:2014p34725,Tilvi:2014p35476}.

Currently, progress is limited by the available near infrared (NIR) spectroscopy at
$z>6$ and the paucity of sources with confirmed \lya\ emission at
$z\gtrsim7$.  Many efforts are underway to increase the spectroscopic
samples 
\citep[][]{Vanzella:2009p29479,Vanzella:2014p36508,Vanzella:2014p33637,Pentericci:2011p27723,Pentericci:2014p34725,Caruana:2012p27502,Caruana:2014p32713,Bradac:2012p28826,Treu:2012p12658,Treu:2013p32132,Balestra:2013p35083,Tilvi:2014p35476,Schenker:2014p35145,Faisst:2014p34184,Karman:2014p37057,Oesch:2015p38504,Watson:2015p38799,Zitrin:2015p40322,Hoag:2015p38918},
although progress from the ground is fundamentally limited by the
Earth's atmosphere.

In this paper, we report on a spectroscopic study of 159
photometrically selected galaxies at $z\gtrsim7$ in the first six
fields targeted by the Grism Lens-Amplified Survey from Space
\citep[GLASS; P.I. T. Treu;][]{Schmidt:2014p33661,Treu:2015p36793}. By
combining \emph{HST}'s NIR slitless
spectroscopic capabilities with the power of the gravitational
magnification by foreground massive galaxy clusters, we carry out the
largest survey of \lya\ emission at $z\gtrsim7$ to date.
We reach 1$\sigma$ line sensitivities of order $5\times10^{-18}$erg/s/cm$^{2}$ over the
wavelength range $0.8-1.7\mu$m, uninterrupted by sky emission or
absorption. 
Including the lensing magnification, $\mu$, of the individual sources, these sensitivities  
improve by a factor of $\mu$, to intrinsic depths which are unreachable without the 
lensing of the foreground clusters. 
Hence, as will become clear in the following, GLASS is providing a unique 
view of the intrinsically fainter emitters, complementary to the bright 
spectroscopically confirmed \lya\ emitters recently presented 
by \cite{Oesch:2015p38504,RobertsBorsani:2015p39891} and \cite{Zitrin:2015p40322}.
We introduce human-based and automated procedures to
identify and quantify the significance of the lines and estimate the
purity and completeness of the sample. After correcting our
statistics for incompleteness and impurity we compare them with
predictions of simple phenomenological models of the \lya\ emission evolution. 
We stack the detections to obtain the first constraint
on the spatial distribution of \lya\ at these redshifts, as well as
limits on the
\lya/CIV and \lya/CIII] line ratios.

The paper is organized as follows. In Section~\ref{sec:glass} we
briefly summarize the GLASS dataset. In Section~\ref{sec:sel} we
introduce our photometric selections and the GLASS grism spectroscopy
of sources at $z\gtrsim7$.  
In Section~\ref{sec:flimandEW}, \ref{sec:pline}, and \ref{sec:stats} we describe
the measurement of flux and equivalent widths of the features
identified as \lya, and estimate the sample completeness and purity. 
In Section~\ref{sec:individualobj} we describe a few interesting cases in detail,
and discuss the implications these could lead to in Section~\ref{sec:ELimplication}.
In Sections~\ref{sec:stack} and \ref{sec:spatialextent} we stack the most
convincing line emitters to look for CIV and CIII] emission,
estimate the spatial extent of \lya\ at $\langle z\rangle=7.2$ and compare it
with simulated $z\sim7.2$ galaxies from the LARS sample, 
before we conclude our study in Section~\ref{sec:conc}.

AB magnitudes \citep{Oke:1974p41298,Oke:1983p41263} and a standard concordance
cosmology with $\Omega_m=0.3$, $\Omega_\Lambda=0.7$, and $h=0.7$
are adopted throughout the paper.

\section{The GLASS Data and Data Reduction}
\label{sec:glass}

GLASS is a 140 orbit slitless spectroscopic survey with \emph{HST}
observing 10 massive galaxy clusters including the 6 Hubble Frontier
Fields clusters (P.I. J. Lotz, HFF) and 8 of the CLASH clusters
\citep[P.I. M. Postman;][]{Postman:2012p27556}.  Taking advantage of
the gravitational lensing of the GLASS clusters, the GLASS grism
spectroscopy reaches flux limits of background sources otherwise
unreachable with the same exposure time.  
An overview of GLASS and its science drivers is given in
the first paper of this series \citep{Treu:2015p36793}. 
One of the key science drivers of GLASS is to study how and when
galaxies reionized the Universe, taking advantage of this
lens-improved depth and emission line detection limit.  Here we
present the first results of this study.

As part of GLASS the core of each cluster has been observed using the \emph{HST} NIR
WFC3 G102 and G141 grisms.  Each grism exposure is accompanied by a
shallower direct image exposure in F105W or F140W to optimize
alignment and extraction of the reduced grism spectroscopy.  The GLASS
observations are split into two distinct position angles (PAs) roughly 90
degrees apart. 
This is done to minimize the number of objects severely affected by
contaminating flux from neighboring objects, and to improve the
identification of emission lines.
The GLASS data were taken following the observing strategy of the
3D-HST survey. The images and spectra are reduced using an updated version of the 3D-HST pipeline
\citep{Brammer:2012p12977,Momcheva:2015p41803}.  The individual grism
exposures are aligned and combined using the AstroDrizzle software
from the DrizzlePac \citep{Gonzaga:2012p26307} and \verb+tweakreg+.
The grism backgrounds are subtracted using sky images from
\cite{Kummel:2011p33451} and \cite{Brammer:2012p12977}.  The direct
images are sky-subtracted by fitting a 2nd order polynomial to the background.  
After alignment and sky-subtraction the final mosaics are interlaced to a
grid of roughly $0\farcs06 \times 12(22)$\AA{} per pixel for the
G102(G141) grisms.  
Before sky-subtraction and interlacing the individual exposures were checked and corrected for backgrounds affected by the Helium Earth-glow described by \cite{Brammer:2014p34990} \citep[see][for details]{Treu:2015p36793}.

The individual spectra of objects detected by 
\verb+SExtractor+ \citep{Bertin:1996p12964} 
in the direct detection image mosaics,
are then extracted from the grism mosaics, using the information about the grism dispersion properties provided in the grism configuration files.
Flux-contamination from neighboring objects is taken into account when extracting the spectra.
For the current study, we generated direct image segmentation maps using 
combined NIR mosaics, including the ancillary CLASH imaging,
for source detection and alignment.
Note that in this way, by predicting the location of the spectral traces from the grism configuration files based on a detection in the ancillary detection images, it is possible to extract spectra for objects (just) outside the grism field-of-view.

For further information on GLASS we refer the reader to
\cite{Schmidt:2014p33661,Treu:2015p36793} and
\url{http://glass.astro.ucla.edu}.

\section{Sample Selection and Spectroscopy}
\label{sec:sel}

\begin{figure*}
\begin{center}
\includegraphics[width=0.99\textwidth]{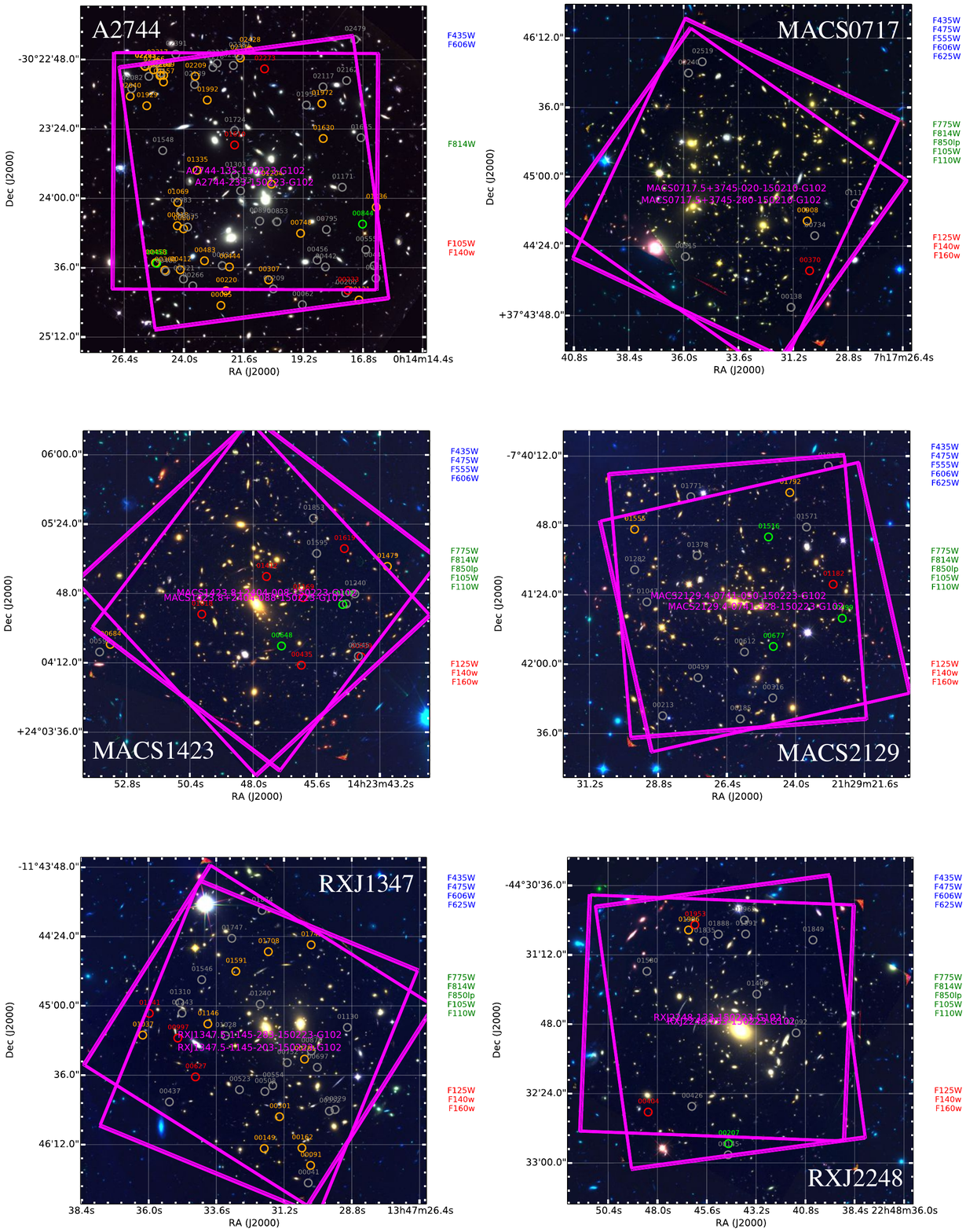}
\caption{\tiny{False-color composite images of the 6 GLASS clusters analyzed in this paper. 
From top left to bottom right we show  \Atwo, \Mseven, \Mfourteen, \Mtwentyone, \Rthirteen, and \Rtwentytwo.
Next to each panel the individual images used to generate the blue, green and red channels of the color composites are listed. 
The magenta polygons mark the field-of-view of the two 90-degrees separated GLASS pointings.
The circles mark the location of the $z\gtrsim7$ objects described in Section~\ref{sec:sel}. 
The orange and gray circles mark the `Gold' and `Silver'  objects from Table~\ref{tab:dropouts}. 
The green and red circles show the location of the `Gold\_EL' and `Silver\_EL' objects presented in Table~\ref{tab:dropouts_EL}. 
The redshift distributions of these sources are shown in Figure~\ref{fig:zhist}.
Note that objects immediately outside the GLASS field-of-view (\Mfourteen{} in the center left panel) can still be recovered and extracted in the grism observations thanks to their detection in the ancillary CLASH imaging.
The apparent over-density of high-redshift objects in \Atwo{} (top left) is caused by the increased depth (compared to the CLASH imaging) of the completed HFF data on \Atwo. A similar improvement in sample size is expected for the remaining 5 HFF clusters in the GLASS sample.}
}
\label{fig:clusters}
\end{center}
\end{figure*}

The sample of high-redshift galaxies analyzed in this study are
selected behind the first 6 completed GLASS clusters \Atwo, \Mseven,
\Mfourteen, \Mtwentyone, \Rthirteen, and \Rtwentytwo.  
We make use of HFF images for A2744, the first HFF cluster with complete GLASS and HFF coverage. 
The remainder of the GLASS/HFF sample will be analyzed and published after the completion of the HFF imaging campaign.
In Figure~\ref{fig:clusters} the color images of these six clusters are shown
with the two 90-degrees separated GLASS pointings indicated by the
magenta polygons.  In the following we describe the photometric
pre-selection of the spectroscopic samples shown by the colored
circles in Figure~\ref{fig:clusters}.

\subsection{Pre-Selection of Spectroscopic Sample}\label{sec:presel}

We assembled an extensive list of Lyman break galaxy candidates at
$z\gtrsim7$ including both existing samples published in the literature and photometric samples selected through multiple color-selections and photometric redshift estimates using the ancillary (NIR-based) CLASH photometry. 
The applied selections and literature samples considered are:
\begin{itemize}
\item[1:] The Lyman break galaxies at $z>7$ investigated by \cite{Zheng:2014p34485}. 

\item[2:] The dropouts and multiple imaged sources presented by \cite{Ishigaki:2015p37005}.

\item[3:] The $z>6$ dropouts and multiple imaged systems presented by \cite{Atek:2014p33154} and \cite{Atek:2015p37010}. 

\item[4-8:] F814W-, F850LP-, F105W-, F110W-, and F125W-dropouts, selected using
the color criteria presented by \cite{Huang:2015p39241}. The selections use
\emph{HST} photometry only. A small subset of candidates have IRAC detections
that support the photometric redshift solutions at $z \gtrsim 7$.
See \cite{Huang:2015p39241} for details.

\item[9:] The components of the geometrically supported redshift 10 candidate multiply imaged system presented by  \cite{Zitrin:2014p36332}

\item[10:] The $z\sim8$ candidate presented by \cite{Laporte:2014p34051}.

\item[11:] The multiply imaged systems from \cite{Lam:2014p35733} above $z=6.5$, i.e., systems 17 and 18.

\item[12:] 
High-redshift candidates from Huang, Hoag, and Brada\v{c} selected as part of follow-up efforts carried out with DEIMOS and MOSFIRE on Keck.

\item[13-14:] 
z- and Y-band dropouts following \cite{Bouwens:2011p8082}, where bands blue-wards of the z/Y band were required to have S/N$<2$.

\item[15:] z-band dropouts selected following \cite{Bouwens:2012p10416}.
Again, bands blue-wards of the z band were all required to have S/N$<2$.

\item[16:] 
JH$_{140}$ dropouts using the criteria described by \cite{Oesch:2013p27877}.
We also searched for YJ and J$_{125}$  dropouts following \cite{Oesch:2013p27877}, but none of these candidates passed our visual inspections, and were therefore not included in any of our final samples.

\item[17:] 
A slightly modified (using F105W instead of F098M)
version of the BoRG $z\sim8$ Y-band dropout selection
\citep{Trenti:2011p12656,Bradley:2012p23263,Schmidt:2014p34189}.

\item[18:]
Galaxies with photometric redshifts $z_\textrm{phot}>6.5$ estimated with the photometric redshift code EA$z$Y \citep{Brammer:2008p13280} run on the CLASH photometry of the CLASH clusters in the sample (all but \Atwo).

\item[19:] 
The CLASH SED-selected $z\gtrsim7$ Lyman break galaxies from \cite{Bradley:2014p32053}.

\item[20-21:] Conservative Photometric selections based on the CLASH  
F850LP, F110W, F125W and F160W photometry. 
All objects from the photometric selections were visually inspected to weed out contaminants and secure clean non-detections in bands blue-wards of F850LP.
\end{itemize}

To summarize, selections 1-3, 9-11, and 19 are all taken from the literature. 
The images of all objects passing the color and spectral energy distribution selections applied to the ancillary photometry by our team (selections 4-8, 12-18, and 20-21) were visually inspected to remove hot pixels, diffraction spikes, and edge defects from the samples. 
We have tabulated this summary in Table~\ref{tab:selectionsummary}.

\tabletypesize{\scriptsize} \tabcolsep=0.15cm
\begin{deluxetable}{llll} \tablecolumns{4}
\tablewidth{0pt}
\tablecaption{Summary of Photometric Pre-Selections\\ of Spectroscopic Sample}
\tablehead{
 \colhead{Cluster} & \colhead{Selection:} & \colhead{Selection:} & \colhead{N$_\textrm{tot}$} \\
  \colhead{} & \colhead{GLASS Team} & \colhead{Literature} & \colhead{} 
}
A2744      			& 4,5,6,7,8						& 1,2,3,9,10,11        		&	11 \\
MACS0717		& 4,5,6,7,8,12,13,14,15,16,17,18		&      	19					& 	13 \\
MACS1423		& 4,5,6,7,8,13,14,15,16,17,18			&	\nodata				& 	11 \\
MACS2129 		& 4,5,6,7,8,13,14,15, 16,17,18   		&	19					& 	12 \\
RXJ1437		& 4,5,6,7,8,13,14,15,16,17,18,20,21		&  	19 					& 	14 \\
RXJ2248		& 4,5,6,7,8,13,14,15,16,17,18	         	&	19					& 	12 \\
\enddata
\label{tab:selectionsummary}
\end{deluxetable}

We split the photometric samples into a `Gold' and `Silver' sample according to the number of times each object was selected. 
Our Gold sample consist of objects picked up by 2 or more of the above selections.
The Gold and Silver samples were furthermore split in an emission line (`EL') and non-emission line sample as described in Section~\ref{sec:specsamp} below. 

The apparent over-density of high-redshift objects in \Atwo{} seen in Figure~\ref{fig:clusters} is caused by the increased depth of the HFF imaging on \Atwo{} compared to the CLASH mosaics, and the extra attention on \Atwo{} this has caused. 
A similar improvement in sample size is expected for the remaining 5 HFF clusters in the GLASS sample, when their completed HFF photometry is available.
We will present these samples in a future publication, when all HFF data will be available on the GLASS clusters.

The final samples of objects are listed in 
Tables~\ref{tab:dropouts} and \ref{tab:dropouts_EL}. The `N$_\textrm{Sel.}/N_\textrm{tot}$', `Sel.', and `$z_\textrm{Sel.}$' columns list the number of selections finding a given object out of the total $N$ selections from Table~\ref{tab:selectionsummary}, which selections include the object, and the mean redshift of the selection(s), respectively.
The `Sample' column lists what sample the objects belong to.

Note that the photometric selections described in this section, should \emph{not} be treated as truly independent selections, as they are all based on essentially the same data, very similar photometry (if not identical) and overlapping selection regions in color space probing the Lyman break, which is also what the photometric redshift selections are sensitive to when searching for high redshift galaxies.

\begin{turnpage}

\tabletypesize{\tiny} \tabcolsep=0.11cm
\begin{deluxetable*}{cccrrcccccccc} \tablecolumns{13}
\tablewidth{0pt}
\tablecaption{$z\gtrsim7$ Dropout Samples with no \lya\ Detection from Visual Inspections}
\tablehead{
 \colhead{Cluster} & \colhead{ID} & \colhead{ID} & \colhead{R.A.} & \colhead{Dec.} & \colhead{P.A.} & \colhead{Sample} & \colhead{N$_\textrm{sel}$/N$_\textrm{tot}$} & \colhead{Sel.} & \colhead{$z_\textrm{Sel.}$} & \colhead{F140W}   &  \colhead{$f_{1\sigma\textrm{limit}}$} & \colhead{$\mu$} \\
 \colhead{} & \colhead{GLASS} & \colhead{Ancillary} & \colhead{[deg]} & \colhead{[deg]} & \colhead{[deg]} & \colhead{} & \colhead{} & \colhead{} & \colhead{} & \colhead{[ABmag]}   &  \colhead{[1e-17 erg/s/cm$^2$]} & \colhead{}
}
 A2744 & 00085 & 03230 & 3.593803625 & -30.415444323 & 135, 233 & Gold & 4 / 11 & 2, 3, 4 & 6.55 & 26.08$\pm$0.05   &                                                                      \nodata      & 3.7$\pm$1.8 \\ 
 A2744 & 00131 & 03158 & 3.570658150 & -30.414663281 & 135, 233 & Gold & 3 / 11 & 2, 3, 4 & 6.25 & 26.62$\pm$0.07  &                                                                       \nodata      & 1.6$\pm$0.4 \\ 
 A2744 & 00220 & 03040 & 3.592948356 & -30.413331741 & 135, 233 & Gold & 2 / 11 & 2, 3 & 5.96 & 27.74$\pm$0.08 &	                                                                         0.94      & 6.6$\pm$4.1 \\ 
 A2744 & 00307 & 02873 & 3.585805956 & -30.411751960 & 135, 233 & Gold & 2 / 11 & 2, 3 & 7.25 & 26.61$\pm$0.04 &	                                                                         0.48      & 3.8$\pm$1.5 \\ 
 A2744 & 00360$^\star$ & 02721 & 3.603208705 & -30.410356491 & 135, 233 & Gold & 4 / 11 & 1, 2, 3, 4 & 6.5 & 27.08$\pm$0.05   &                                                                      0.54      & 3.7$\pm$7.5 \\ 
 A2744 & 00412 & 02732 & 3.600611950 & -30.410302069 & 135, 233 & Gold & 3 / 11 & 2, 3, 4 & 6.40 & 28.29$\pm$0.19    &                                                                     0.57      & 9.2$\pm$3.4 \\ 
 A2744 & 00444$^\star$ & 02676 & 3.592367074 & -30.409889954 & 135, 233 & Gold & 4 / 11 & 1, 2, 3, 11 & 7.39 & 28.86$\pm$0.12   &                                                                      0.35      & 7.0$\pm$7.1 \\ 
 A2744 & 00458 & 02627 & 3.604762132 & -30.409290304 & 135, 233 & Gold & 2 / 11 & 3, 4 & 6.53 & 27.80$\pm$0.07&	                                                                         0.52      & 2.9$\pm$8.5 \\ 
 A2744 & 00483 & 02686 & 3.596557317 & -30.409003929 & 135, 233 & Gold & 2 / 11 & 2, 3 & 7.25 & 27.13$\pm$0.07   &                                                                       0.36      & 5.0$\pm$3.4 \\ 
 A2744 & 00748 & 02234 & 3.580452097 & -30.405043370 & 135, 233 & Gold & 3 / 11 & 2, 3, 11 & 6.96 & 26.94$\pm$0.06   &                                                                      0.40      & 5.6$\pm$1.1 \\ 
 A2744 & 00807 & 02178 & 3.600055342 & -30.404393062 & 135, 233 & Gold & 2 / 11 & 2, 3 & 7 & 27.18$\pm$0.07  &                                                                        0.39      & 4.8$\pm$3.4 \\ 
 A2744 & 00818 & 02135 & 3.601100197 & -30.403956945 & 135, 233 & Gold & 3 / 11 & 2, 3, 4 & 6.25 & 27.90$\pm$0.07 &                                                                        0.66      & 3.5$\pm$1.4 \\ 
 A2744 & 01036 & 01942 & 3.567777944 & -30.401277987 & 135, 233 & Gold & 2 / 11 & 3, 4 & 6.46 & 27.39$\pm$0.16   &                                                                       0.56      & 2.1$\pm$0.9 \\ 
 A2744 & 01069 & 01891 & 3.601044487 & -30.400590602 & 135, 233 & Gold & 2 / 11 & 1, 3 & 7.45 & 27.00$\pm$0.06  &                                                                        0.34      & 2.8$\pm$0.8 \\ 
 A2744 & 01204$\dagger$ & -00088 & 3.585323923 & -30.397960001 & 135, 233 & Gold & 3 / 11 & 2, 3, 11 & 6.90 & 27.16$\pm$0.07$^\dagger$   &                                              0.44      & 3.2$\pm$2.8 \\ 
 A2744 & 01335$^\star$ & 01506 & 3.597814977 & -30.395957621 & 135, 233 & Gold & 3 / 11 & 2, 3, 11 & 7 & 26.58$\pm$0.04   &                                                                       0.39      & 2.9$\pm$0.9 \\ 
 A2744 & 01929 & 00847 & 3.606221824 & -30.386645344 & 135, 233 & Gold & 2 / 11 & 2, 3 & 5.80 & 25.98$\pm$0.03   &                                                                       1.13      & 1.7$\pm$0.7 \\ 
 A2744 & 01972 & 00816 & 3.576890999 & -30.386328547 & 135, 233 & Gold & 3 / 11 & 2, 3, 4 & 6.44 & 28.22$\pm$0.11   &                                                                       0.57      & 4.4$\pm$7.6 \\ 
 A2744 & 01992$^\star$ & 00765 & 3.596089446 & -30.385830967 & 135, 233 & Gold & 4 / 11 & 2, 1, 3, 6 & 8 & 26.54$\pm$0.04  &                                                                        0.30      & 2.5$\pm$5.6 \\ 
 A2744 & 02040 & 00723 & 3.608995192 & -30.385282140 & 135, 233 & Gold & 3 / 11 & 2, 3, 4 & 6.10 & 27.97$\pm$0.21   &                                                                       1.61      & 1.5$\pm$1.0 \\ 
 A2744 & 02157 & 00557 & 3.603418234 & -30.383215863 & 135, 233 & Gold & 3 / 11 & 2, 3, 4 & 5.80 & 27.69$\pm$0.09    &                                                                      1.16      & 1.7$\pm$0.9 \\ 
 A2744 & 02193 & 00477 & 3.603853194 & -30.382264279 & 135, 233 & Gold & 3 / 11 & 1, 2, 3 & 8.40 & 26.91$\pm$0.04   &                                                                          0.44      & 1.6$\pm$0.9 \\ 
 A2744 & 02199$^\star$ & 00469 & 3.603383290 & -30.382256248 & 135, 233 & Gold & 4 / 11 & 1, 2, 3, 6 & 8.10 & 25.82$\pm$0.04  &                                                                           0.44      & 1.6$\pm$0.9 \\ 
 A2744 & 02204 & 00479 & 3.604003006 & -30.382306486 & 135, 233 & Gold & 2 / 11 & 1, 2 & 8.10 & 27.74$\pm$0.07   &                                                                          0.44      & 1.6$\pm$0.9 \\ 
 A2744 & 02209 & 00487 & 3.598091105 & -30.382391542 & 135, 233 & Gold & 2 / 11 & 1, 3 & 7.64 & 27.79$\pm$0.16    &                                                                         0.31      & 1.8$\pm$2.3 \\ 
 A2744 & 02266 & 00433 & 3.605063809 & -30.381462296 & 135, 233 & Gold & 3 / 11 & 1, 2, 3 & 7.70 & 27.99$\pm$0.14 &                                                                            0.47      & 1.5$\pm$0.8 \\ 
 A2744 & 02283$^\star$ & 00600 & 3.606467680 & -30.380994116 & 135, 233 & Gold & 4 / 11 & 1, 2, 3, 6 & 7.80 & 27.09$\pm$0.04  &                                                                           0.45      & 1.5$\pm$1.0 \\ 
 A2744 & 02295 & 00599 & 3.606564953 & -30.380917190 & 135, 233 & Gold & 3 / 11 & 1, 2, 3 & 7.60 & 27.07$\pm$0.04  &                                                                           0.46      & 1.5$\pm$1.0 \\ 
 A2744 & 02317 & 00333 & 3.604519959 & -30.380466741 & 135, 233 & Gold & 5 / 11 & 1, 2, 3, 6, 10 & 8 & 25.86$\pm$0.04 &                                                                            0.44      & 1.5$\pm$0.8 \\ 
 A2744 & 02379 & 00265 & 3.590532446 & -30.379764602 & 135, 233 & Gold & 2 / 11 & 2, 3 & 6.10 & 27.97$\pm$0.10  &                                                                           0.76      & 2.2$\pm$1.6 \\ 
 A2744 & 02428 & 00163 & 3.588984152 & -30.378668677 & 135, 233 & Gold & 3  / 11& 1, 2, 3 & 7.89 & 27.87$\pm$0.07  &                                                                           0.44      & 2.1$\pm$1.6 \\ 
 MACS0717 & 00908 & 01656 & 109.377446020 & 37.743640029 & 020, 280 & Gold & 2 / 13 & 12, 15 & 7.25 & 27.25$\pm$0.19 &                                                                         0.44      & 8.5$\pm$19.3 \\ 
 MACS1423 & 00684 & 01408 & 215.972592500 & 24.072659477 & 008, 088 & Gold & 3 / 11 & 13, 15, 17 & 7 & 27.27$\pm$0.19 &                                                                            0.52      & \nodata \\ 
 MACS1423 & 01479 & 00656 & 215.928811200 & 24.083905686 & 008, 088 & Gold & 3 / 12 & 11, 15, 17 & 7 & 26.14$\pm$0.11 &                                                                            0.68      & \nodata \\ 
 MACS2129 & 01555 & 00475 & 322.373418740 & -7.680549573 & 050, 328 & Gold & 2 / 12 & 13, 15 & 7 & 27.87$\pm$0.27 &                                                                        0.44      & \nodata \\ 
 MACS2129 & 01792 & 00218 & 322.350848970 & -7.675244331 & 050, 328 & Gold & 2 / 12 & 18, 19 & 6.85 & 27.25$\pm$0.19 &                                                                            0.12      & \nodata \\ 
 RXJ1347 & 00091 & 02025 & 206.876076960 & -11.772996916 & 203, 283 & Gold & 2 / 14 & 18, 19 & 6.82 & 27.70$\pm$0.26 &                                                                         0.78      & \nodata \\ 
 RXJ1347 & 00149 & 01954 & 206.882922680 & -11.770563897 & 203, 283 & Gold & 2 / 14 & 5, 21 & 7 & 26.18$\pm$0.11 &                                                                            0.45      & \nodata \\ 
 RXJ1347 & 00162 & 01951 & 206.877358800 & -11.770481642 & 203, 283 & Gold & 3 / 14 & 13, 15, 20 & 7 & 28.33$\pm$0.39  &                                                                           0.27      & \nodata \\ 
 RXJ1347 & 00301 & 01777 & 206.880670900 & -11.765976036 & 203, 283 & Gold & 2 / 14 & 13, 15 & 7 & 26.42$\pm$0.14 &                                                                            0.46      & \nodata \\ 
 RXJ1347 & 00781 & 01316 & 206.876976150 & -11.757678122 & 203, 283 & Gold & 4 / 14 & 14, 17, 18, 19 & 7.5 & 26.97$\pm$0.16  &                                                                           0.38      & \nodata \\ 
 RXJ1347 & 01037$^\diamond$ & 01046 & 206.900859670 & -11.754209621 & 203, 283 & Gold & 4 / 14 & 5, 18, 19, 21 & 7 & 26.09$\pm$0.09 &                                                                            0.43      & \nodata \\ 
 RXJ1347 & 01146 & 00943 & 206.891246090 & -11.752606761 & 203, 283 & Gold & 4 / 14 & 6, 19, 20, 21 & 7.5 & 26.38$\pm$0.12  &                                                                         0.38      & \nodata \\ 
 RXJ1347 & 01591 & 00471 & 206.887118070 & -11.745016973 & 203, 283 & Gold & 2 / 14 & 20, 21 & 7 & 25.65$\pm$0.07 &                                                                            0.44      & \nodata \\ 
 RXJ1347 & 01708 & 00346 & 206.882316460 & -11.742182707 & 203, 283 & Gold & 5  / 14& 13, 15, 17, 19, 20 & 7 & 26.59$\pm$0.13  &                                                                           0.45      & \nodata \\ 
 RXJ1347 & 01745 & 00310 & 206.876001920 & -11.741194080 & 203, 283 & Gold & 3 / 14 & 18, 19, 20 & 7 & 26.98$\pm$0.17 &                                                                            0.44      & \nodata \\ 
 RXJ2248 & 01906 & 00253 & 342.193691160 & -44.516422494 & 053, 133 & Gold & 2 / 12 & 14, 17 & 8 & 27.60$\pm$0.25  &                                                                           0.47      & 5.5$\pm$3.0 \\ 
\hline
 A2744 & 00431 & 02609 & 3.593576781 & -30.409700762 & 135, 233 & Silver & 1 / 11 & 3 & 6.75 & 26.78$\pm$0.04   &                                                                      0.46      & 5.3$\pm$1.9 \\   
 A2744 & 00795 & 02186 & 3.576122532 & -30.404490552 & 135, 233 & Silver & 1 / 11 & 3 & 6.75 & 26.61$\pm$0.05  &                                                                        0.48      & 3.5$\pm$2.0 \\  
\nodata & \nodata & \nodata & \nodata & \nodata & \nodata & \nodata & \nodata & \nodata & \nodata & \nodata &\nodata & \nodata 
\enddata
\tablecomments{
The `Cluster' column lists the cluster the objects was found in.
`ID GLASS' designates the ID of the object in the GLASS detection catalogs. Note that these IDs are \emph{not} identical to the IDs of the v001 data releases available at \url{https://archive.stsci.edu/prepds/glass/} presented by \cite{Treu:2015p36793}, as a more aggressive detection threshold and de-blending scheme was used for the current study. 
`ID Ancillary' lists the ids from the ancillary A2744 HFF+GLASS and CLASH IR-based photometric catalogs. 
`R.A.' and `Dec.' list the J2000 coordinates of each object.
`P.A.' lists the position angle of the two GLASS orientations (the PA\_V3 keyword of image fits header). 
The `Sample' column indicates what sample the object belongs to.
`N$_\textrm{sel}$/N$_\textrm{tot}$' list the number of photometric selections picking out each object and the total number of selections applied to the data set from Table~\ref{tab:selectionsummary}. 
The actual selections listed in the `Sel.' column are described in Section~\ref{sec:sel}. 
The $z_\textrm{Sel.}$ column lists the median redshift of the $N_\textrm{Sel.}$ selections containing the object. 
`F140W' lists the AB magnitude of the objects.
The $f_{1\sigma\textrm{limit}}$ column quotes the line flux limit for the emission lines obtained as described in Section~\ref{sec:flimandEW}.
The $\mu$ column gives the magnifications of the HFF clusters obtained as described in Section~\ref{sec:flimandEW}.
The complete Silver sample is available upon request.
$^\diamond$RXJ1347\_01037 has a confirmed redshift from the GLASS spectra and from Keck DEIMOS as described in Section~\ref{sec:rxj1347_1037}. 
Its GLASS spectra are shown in Figure~\ref{fig:z7obj}.
$^\dagger$Object had no good counterpart ($r_\textrm{match}>1\farcs0$) in the default photometric catalog, so its magnitude comes from a more aggressive (with respect to de-blending and detection threshold) re-run of SExtractor.
$^\star$Objects searched for CIII]$\lambda1909$ by \cite{Zitrin:2015p39322} as described in the text. 
}
\label{tab:dropouts}
\end{deluxetable*}
    
\end{turnpage}

\begin{turnpage}

\tabletypesize{\tiny} \tabcolsep=0.11cm
\begin{deluxetable*}{cccrrcccccccccc} \tablecolumns{15}
\tablewidth{0pt}
\tablecaption{$z\gtrsim7$ Dropout Samples with \lya\ Detections from Visual Inspection}
\tablehead{
 \colhead{Cluster} & \colhead{ID} & \colhead{ID} & \colhead{R.A.} & \colhead{Dec.} & \colhead{P.A.} & \colhead{Sample} & \colhead{N$_\textrm{sel}$/N$_\textrm{tot}$} & \colhead{Sel.} & \colhead{$z_\textrm{Sel.}$ ($z_\textrm{\lya}$)} & \colhead{F140W} & \colhead{$\lambda_\textrm{lines}$}  & \colhead{EW$_\textrm{\lya}$} & \colhead{$f_\textrm{line}$ or $f_{1\sigma\textrm{limit}}$} & \colhead{$\mu$} \\
 \colhead{} & \colhead{GLASS} & \colhead{Ancillary} & \colhead{[deg]} & \colhead{[deg]} & \colhead{[deg]} & \colhead{} & \colhead{} & \colhead{} & \colhead{} & \colhead{[ABmag]} & \colhead{$\pm$50\AA}  & \colhead{[\AA]} & \colhead{[1e-17 erg/s/cm$^2$]} & \colhead{}
}
\startdata 
 A2744 & 00463 & 02720 	& 3.604573038 			& -30.409357092 & 135, 233 & Gold\_EL 	& 3 / 11 	& 2, 3, 4 	& 6.1	(6.73)  & 28.19$\pm$0.13 				& 9395, \nodata &                               379$\pm$147,  \nodata 		& 1.91$\pm$0.7,  0.65		    	& 2.9$\pm$7.3 \\ 
 A2744 & 00844$^\circ$ & 02111 	& 3.570068923 			& -30.403715689 & 135, 233 & Gold\_EL 	& 2 / 11 	& 3, 4 		& 6.5	(6.34)  & 26.85$\pm$0.04 				& 8929, \nodata &                               152$\pm$52,  \nodata 		& 2.76$\pm$0.94,  0.88    		& 2.0$\pm$0.8 \\ 
 MACS1423 & 00648 & 01418 	& 215.945534620 			& 24.072435174 & 008, 088 & Gold\_EL 	& 3 / 11 	& 13, 15, 17 	& 7	(6.88)  & 26.05$\pm$0.09 				& 9585, \nodata &                                 56$\pm$16,  \nodata 		& 2.0$\pm$0.56,  0.71    		& \nodata \\ 
 MACS1423 & 01102$^\diamond$ & 01022 	& 215.935869430 	& 24.078415134 & 008, 088 & Gold\_EL 	& 2  / 11	& 5, 13 			& 7	(6.96)  & 26.56$\pm$0.12 	& 9681, \nodata &	         	       85$\pm$30, \nodata  		& 1.87$\pm$0.63, 1.04     		& \nodata \\ 
 MACS2129 & 00677$^\diamond$ & 01408 	& 322.353239440 	& -7.697441500 & 050, 328 & Gold\_EL 	& 4 / 12 	& 13, 15, 17, 19 	& 7	(6.88)  & 27.17$\pm$0.17 	& 9582, 9582 &                               272$\pm$80,  270$\pm$70 		& 3.45$\pm$0.87,  3.42$\pm$0.70    	& \nodata \\ 
 MACS2129 & 00899$^\dagger$ & 01188 	& 322.343220360 	& -7.693382243 & 050, 328 & Gold\_EL 	& 2 / 12 	& 7, 14 			& 8.5	(8.10)  & 26.69$\pm$0.13 	& 11059, 11069 &                            44$\pm$31,  74$\pm$29 		& 0.74$\pm$0.52,  1.26$\pm$0.47   	& \nodata \\ 
 MACS2129 & 01516 & 00526 	& 322.353942530 			& -7.681646419 & 050, 328 & Gold\_EL 	& 2 / 12 	& 13, 15 	& 7	(6.89)  & 28.41$\pm$0.33 				& 9593, \nodata &                               668$\pm$290,  \nodata 		& 2.7$\pm$0.85,  0.58    		& \nodata \\ 
 RXJ2248 & 00207 & 01735 	& 342.185601570 			& -44.547224418 & 053, 133 & Gold\_EL 	& 2 / 12 	& 13, 15 	& 7	(8.55)  & 28.61$\pm$0.45 				& 11609, \nodata &                              920$\pm$543,  \nodata	 	& 2.55$\pm$1.07,  \nodata    		& 2.1$\pm$0.8 \\ 
\hline
 A2744 & 00233 & 03032 	& 3.572513845 			& -30.413266331 & 135, 233 & Silver\_EL 	& 1 / 11 & 1 			& 8.5 	(8.17)     & 28.36$\pm$0.18 				& 11156, \nodata &                        804$\pm$338,  \nodata	 	& 2.95$\pm$1.14,  0.74    		& 1.9$\pm$0.4 \\ 
 A2744 & 01610 & 01282 	& 3.591507273 			& -30.392303082 & 135, 233 & Silver\_EL 	& 1 / 11 & 4 			& 6.5	 (5.91)     & 27.83$\pm$0.07 				& \nodata, 8406 &                              \nodata,  355$\pm$178 		& 1.35,  2.79$\pm$1.39    		& 8.6$\pm$27.0 \\ 
 A2744 & 02273 & 00420 	& 3.586488763 			& -30.381334667 & 135, 233 & Silver\_EL 	& 1 / 11 & 3 			& 5.71	(6.17)     & 28.48$\pm$0.12 				& 8717, \nodata &                              766$\pm$257,  \nodata 		& 3.21$\pm$1.01,  1.02    		& 2.7$\pm$6.6 \\ 
 MACS0717 & 00370 & 02063 	& 109.377007840 			& 37.736462661 & 020, 280 & Silver\_EL 	& 1 / 13 & 14 			& 7.5	(6.51)     & 27.66$\pm$0.28 				& 9138, \nodata &                              221$\pm$102,  \nodata 		& 1.87$\pm$0.72,  0.43    		& 2.5$\pm$1.4 \\ 
 MACS1423 & 00435 & 01567 	& 215.942403590			& 24.069659639 & 008, 088 & Silver\_EL 	& 1 / 11 & 18 			& 7.27	(7.63  )   & 25.29$\pm$0.06 				& 10500, \nodata &                             15$\pm$7,  \nodata 		& 1.01$\pm$0.47,  0.54    		& \nodata \\ 
 MACS1423 & 00539 & 01526 	& 215.932958480 			& 24.070875663 & 008, 088 & Silver\_EL 	& 1 / 11 & 5 			& 7	(6.13)     & 25.99$\pm$0.09 				& 8666, \nodata &                             89$\pm$29,  \nodata 		& 3.7$\pm$1.17,  0.75    		& \nodata \\ 
 MACS1423 & 01018$^\star$ & 01128 	& 215.958132710 		& 24.077013896 & 008, 088 & Silver\_EL 	& 1 / 11 & 14 			& 8	(10.27)   & 27.81$\pm$0.21 				& 13702, \nodata &                              558$\pm$176,  \nodata 		& 2.72$\pm$0.67,  0.42    		& \nodata \\ 
 MACS1423 & 01169 & 00954 	& 215.942112130 			& 24.079404012 & 008, 088 & Silver\_EL 	& 1 / 11 & 6 			& 8	(6.99)     & 26.01$\pm$0.10 				& 9721, \nodata &                                62$\pm$18,  \nodata 		& 2.26$\pm$0.61,  0.68    		& \nodata \\ 
 MACS1423 & 01412 & 00756 	& 215.947908420 			& 24.082450925 & 008, 088 & Silver\_EL 	& 1 / 11 & 15 			& 7	(6.77)   	  & 27.84$\pm$0.24 				& 9448, \nodata &                            190$\pm$82,  \nodata 		& 1.31$\pm$0.49,  0.79    		& \nodata \\ 
 MACS1423 & 01619 & 00526 	& 215.935606220 			& 24.086476168 & 008, 088 & Silver\_EL 	& 1 / 11 & 5 			& 7	(7.17)     & 26.53$\pm$0.12 				& 9932, \nodata &                            59$\pm$27,  \nodata 		& 1.31$\pm$0.57,  0.51    		& \nodata \\ 
 MACS2129 & 01182 & 00914 	& 322.344533970 			& -7.688477035 & 050, 328 & Silver\_EL 	& 1 / 12 & 14 			& 8	(8.99)     & 27.64$\pm$0.20 				& 12145, \nodata &                              606$\pm$185,  \nodata 		& 3.92$\pm$0.94,  0.54    		& \nodata \\ 
 RXJ1347 & 00627$^\diamond$ & 01488 	& 206.893075800 	& -11.760237310 & 203, 283 & Silver\_EL 	& 1 / 14 & 13 	& 7	(7.84)     & 27.85$\pm$0.26 				& 10750, \nodata &                              290$\pm$118,  \nodata 		& 1.76$\pm$0.57,  0.45    		& \nodata \\ 
 RXJ1347 & 00997 & 01070 	& 206.895685760 			& -11.754637616 & 203, 283 & Silver\_EL 	& 1 / 14 & 13 			& 7	(6.79)     & 26.94$\pm$0.20 				& 9467, 9463 &                           149$\pm$54,  90$\pm$45 		& 2.37$\pm$0.75,  1.42$\pm$0.66    	& \nodata \\ 
 RXJ1347 & 01241$^\diamond$ & 00864 	& 206.899894840 	& -11.751082858 & 203, 283 & Silver\_EL 	& 1 / 14 & 5 	& 7	(7.14)     & 26.68$\pm$0.16 				& 9902, \nodata &                         522$\pm$87,  \nodata 		& 10.05$\pm$0.84,  0.54    		& \nodata \\ 
 RXJ2248 & 00404$^\star$ 	& 01561 	& 342.201879400 	& -44.542663866 & 053, 133 & Silver\_EL 	& 1 / 12 & 7 			& 9	(9.89)     & 27.05$\pm$0.18 				& 13239, \nodata &                               142$\pm$64,  \nodata 		& 1.45$\pm$0.61,  0.41    		& 1.8$\pm$0.5 \\ 
 RXJ2248 & 01953 & 00220 	& 342.192399500 			& -44.515663484 & 053, 133 & Silver\_EL 	& 1 / 12 & 14 			& 8	(6.50)     & 27.99$\pm$0.27 				& \nodata, 9118 &                               \nodata,  686$\pm$227 		& 0.95,  4.3$\pm$0.94    		& 3.8$\pm$1.9  
\enddata
\tablecomments{
The `Cluster' column lists the cluster the objects was found in.
`ID GLASS' designates the ID of the object in the GLASS detection catalogs. Note that these IDs are \emph{not} identical to the IDs of the v001 data releases available at \url{https://archive.stsci.edu/prepds/glass/} presented by \cite{Treu:2015p36793}, as a more aggressive detection threshold and de-blending scheme was used for the current study. 
`ID Ancillary' lists the ids from the ancillary A2744 HFF+GLASS and CLASH IR-based photometric catalogs. 
`R.A.' and `Dec.' list the J2000 coordinates of each object.
`P.A.' lists the position angle of the two GLASS orientations (the PA\_V3 keyword of image fits header). 
The `Sample' column indicates what sample the object belongs to.
`N$_\textrm{sel}$/N$_\textrm{tot}$' list the number of photometric selections picking out each object and the total number of selections applied to the data set from Table~\ref{tab:selectionsummary}. 
The actual selections listed in the `Sel.' column are described in Section~\ref{sec:sel}. 
The $z_\textrm{Sel.}$ column lists the median redshift of the $N_\textrm{Sel.}$ selections containing the object, followed by the \lya\ redshift for the emission line. 
`F140W' lists the AB magnitude of the objects.
The column `$\lambda_\textrm{lines}$' lists the wavelength of the detected emission lines.
The equivalent width of the \lya\ emission lines is given in EW$_\textrm{\lya}$.
$f_\textrm{line}$ and $f_{1\sigma\textrm{limit}}$ quotes the line flux  and the flux limit for the emission lines obtained as described in Section~\ref{sec:flimandEW}.
The $\mu$ column gives the magnifications of the HFF clusters obtained as described in Section~\ref{sec:flimandEW}.
In columns containing two values separated by a comma, the individual values refer to the corresponding PAs of the GLASS data listed in the column PA 
$^\star$As described in Section~\ref{sec:z10} these two objects are potential low-redshift contaminants.
$^\dagger$The particularly interesting redshift 8.1 candidate MACS2129\_00899 is discussed in detail in Section~\ref{sec:macs2129_899}.
$^\diamond$Object's G102 grism spectra at the two GLASS PAs are shown in Figure~\ref{fig:fobjects}.
\textbf{
$^\circ$Object is included in the \cite{Atek:2014p33154} sample. After updating the photometry in \cite{Atek:2015p37010} the object no longer satisfy their selection criteria. Even though, the detected emission line in the GLASS spectra agrees well with the photometric redshift, the fact that updated (optical) photometry dis-regards this object as a $z>6$ source speaks in favor of the object being a contaminating low-redshift line emitter. Spectroscopic follow-up is needed to confirm this.
} 
}
\label{tab:dropouts_EL}
\end{deluxetable*}


\end{turnpage}

\subsection{Purity and Completeness of Photometric Samples}

Photometrically selected samples of high-redshift galaxies are know to be both incomplete and contaminated by low-redshift sources.
The incompleteness is usually a consequence of searching for high-redshift galaxies at the detection limits of the imaging data, and in the low-S/N regime. 
Photometric interlopers and contaminants occur as objects mimick the colors of high-redshift galaxies. 
In particular the rest-frame 4000\AA\ break in star forming galaxies is know to contaminate Lyman break galaxy samples, as the resulting colors from a 4000\AA\ break are very similar to the ones obtained from a Lyman Break.
Also spurious sources and cool dwarf stars are known to mimic the colors of high-redshift galaxies and contaminate Lyman break samples. For detailed discussions on high-redshift galaxy sample contaminants we refer to, e.g., \cite{Dunlop:2013p23759,Coe:2013p26313,Wilkins:2014p33959,Finkelstein:2015p37430} or \cite{Bouwens:2015p34683}.
As is the case for the completeness of high-redshift samples, the contamination is often a results of lacking depth of the photometry, in particular blue-wards of the Lyman break where non-detections are required, as exemplified by a few \emph{HST} sources by \cite{Laporte:2015p37877}.
Estimated contamination fractions in high-redshift samples range from ten percent to over forty \cite[e.g.][]{Malhotra:2005p27597,Stanway:2003p27707,Schmidt:2014p34189}.
\textbf{
As the Silver sample objects were only picked up by one photometric selection, we note that these objects must be considered having a higher risk of being low-$z$ contaminants.
}
Irrespective of the type and cause of the contamination and lack of completeness, when performing inference using photometrically selected high-redshift galaxy samples, both the contaminants, i.e. the purity, and the completeness need to be properly accounted for.
This is often done via visual inspection (to remove contamination from stars and spurious sources) and simulations \cite[e.g.,][]{Oesch:2007p30657,Oesch:2012p30149,Bouwens:2011p8082}. 
In the present study we focus on the \emph{spectroscopic} line emitter samples presented in Section~\ref{sec:specsamp}, and the completeness and purity of the photometric pre-selection does therefore not affect our measurements, given that all the sources have detected line emission.
We do have to worry about contamination by low-redshift line emitters, however,
As we will describe in Section~\ref{sec:individualobj}, emission line galaxy samples are potentially contaminated by, e.g., low-redshift [OII]$\lambda$3727 emitters. 
Broad wavelength coverage to confirm non-detections of other low-$z$ emission lines and high-resolution spectroscopy to resolve line morphology of individual lines can be used to account for this contamination, as we will show in Section~\ref{sec:individualobj}. 

\subsection{Finalizing the Spectroscopic Samples}\label{sec:specsamp}

We extracted the GLASS spectra of all candidates in the Gold and Silver samples detected in the NIR detection image mosaics (cf. Section~\ref{sec:glass}).
Faint photometric candidates from the literature (from high redshift candidate searches including HFF data on \Atwo), that were not detected in our NIR mosaics, where not extracted, and are therefore not included in Tables~\ref{tab:dropouts} and \ref{tab:dropouts_EL}. 
As noted, we will present these sources in a future publication. 

\begin{figure}
\begin{center}
\includegraphics[width=0.45\textwidth]{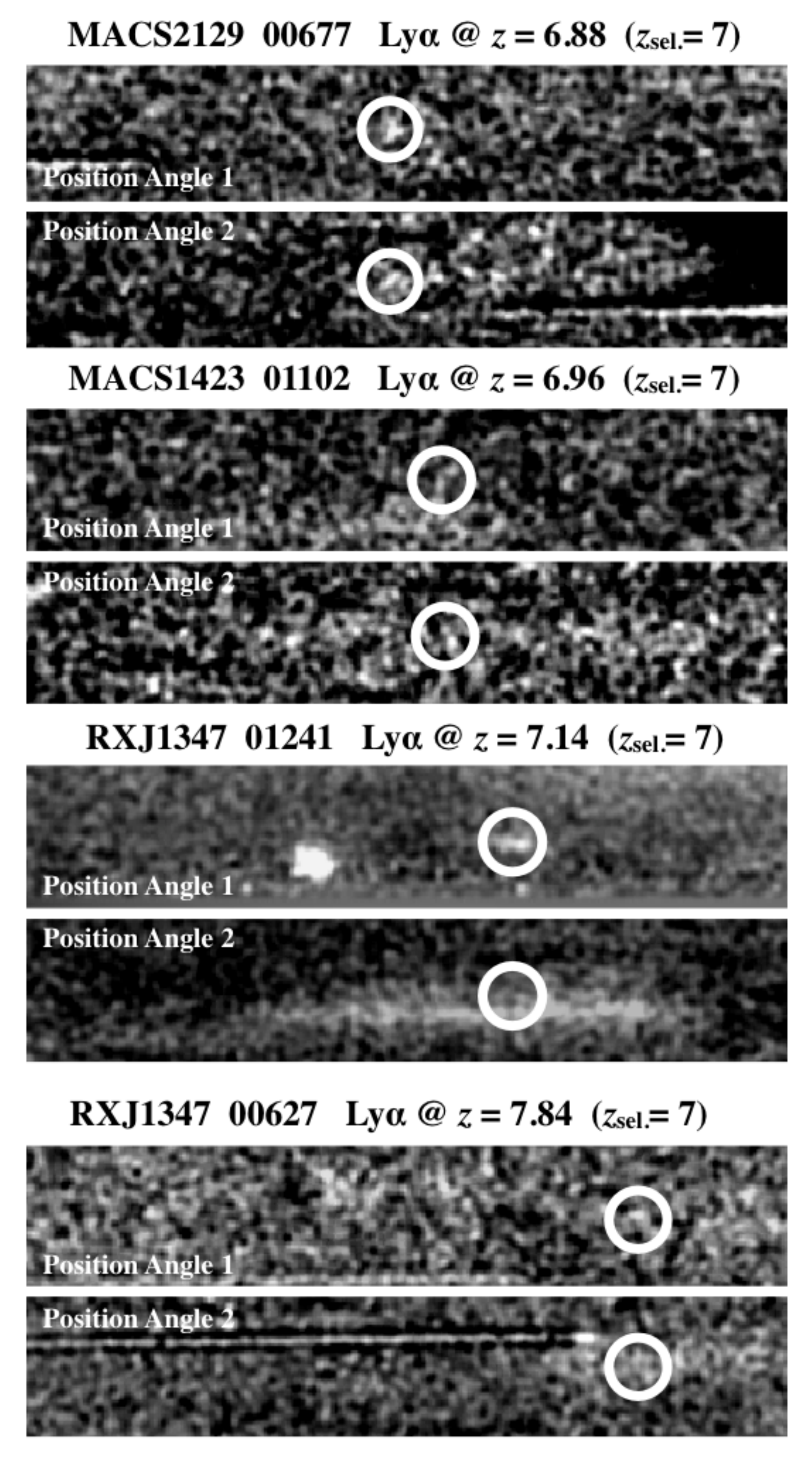}
\caption{Examples of GLASS spectra for 4 out of the 24 $z\gtrsim7$ emission line objects listed in Table~\ref{tab:dropouts_EL}. For each object the G102 spectrum at both of the GLASS PAs are shown. The assumed \lya\ redshift and the selection redshift from Table~\ref{tab:dropouts_EL} are quoted above each panel. The circles mark the location of the emission lines.
All spectra have been subtracted the contamination model from the GLASS reduction.
}
\label{fig:fobjects}
\end{center}
\end{figure}

The extracted GLASS spectra were visually inspected using the publicly available GLASS inspection GUIs \verb+GiG+ and \verb+GiGz+ \citep[see Appendix~A of][and \url{https://github.com/kasperschmidt/GLASSinspectionGUIs}]{Treu:2015p36793} by 3-4 GLASS team members (KBS, TT, MB, BV and LP) to identify emission lines.
The wavelength of any potential (\lya) emission was noted and subsequently compared to the other independent inspections. 
If an emission line was marked by 2 or more inspectors (within $\pm$50\AA) the object was re-inspected by KBS and TT. 
The candidates deemed to be real upon re-inspection constitute the emission line sample presented in Table~\ref{tab:dropouts_EL}. 
In summary, we have assembled a total of 159 unique high-$z$ galaxies with redshifts $\gtrsim7$ and GLASS spectroscopy in the G102 and G141 grisms.
Of these, 55 are found in at least two different pre-selections (Gold) out of which 8 have emission lines consistent with \lya\ (Gold\_EL).
A total of 104 objects were only selected by one pre-selection (Silver). Of these 16 have promising lines consistent with \lya\ (Silver\_EL).
In Figure~\ref{fig:fobjects} we show four examples of emission line objects detected in the GLASS data. 
Each of the four panels shows the spectra from the two distinct GLASS PAs with the location of the emission line marked by the white circles.

As illustrated by the emission line wavelengths listed in Table~\ref{tab:dropouts_EL}, the emission lines were not visually identified in both PAs in the majority of the objects.
If the contamination model was perfect, the exposure times were identical and the background level was constant in the data from the two different PAs, the signal-to-noise (S/N) of any detected lines should be the same in the two GLASS spectra.
However, given the varying background from the Helium Earth-glow mentioned in Section~\ref{sec:glass} (which changes the effective exposure time by up to 13\% between PAs), the change in the contribution to the background from the intra-cluster light in the two dispersion directions, and residuals from subtracting the contamination models (causing larger flux uncertainties and altered background levels), it is not surprising that several of the moderate S/N line detections are only seen in one PA.
We consider objects with lines clearly detected in both PAs like MACS2129 00677 shown in Figure~\ref{fig:fobjects} to be particularly strong line emitter candidates.

To our knowledge the only spectroscopically confirmed object in our sample of $z\gtrsim7$ sources is RXJ1347\_01037 at $z=6.76$ \citep{Huang:2015p39241}, which we will describe in more details in Section~\ref{sec:individualobj}.

At redshifts just below 6.5 (and therefore not included in the samples described here) a few objects have been spectroscopically confirmed.
In Appendix~\ref{sec:VandBB} we describe the two known spectroscopically confirmed multiple imaged systems at $z=6.1$ \citep{Boone:2013p35081,Balestra:2013p35083} and $z=6.4$ \citep{Vanzella:2014p33637}.

\begin{figure*}
\begin{center}
\includegraphics[width=0.49\textwidth]{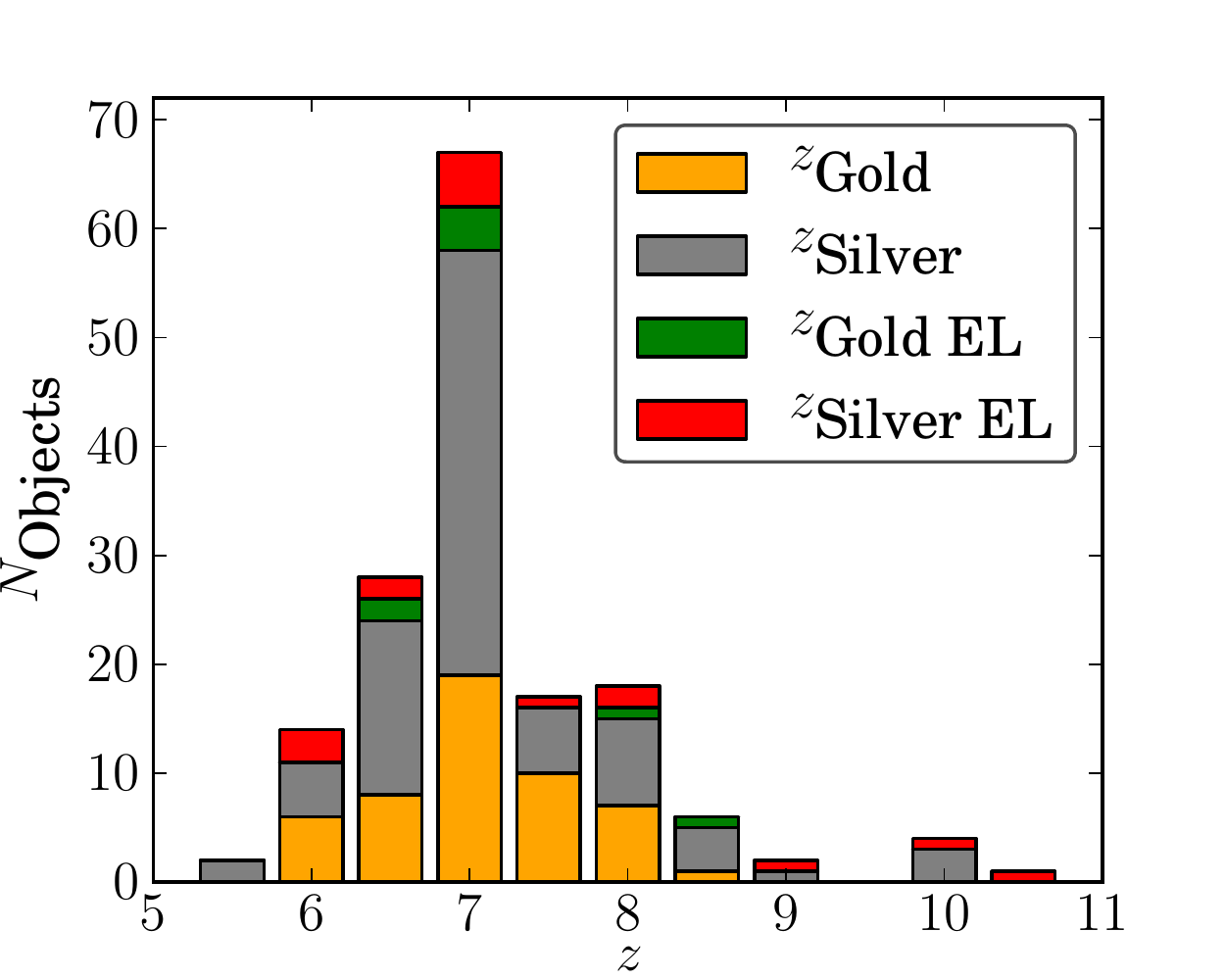}
\includegraphics[width=0.49\textwidth]{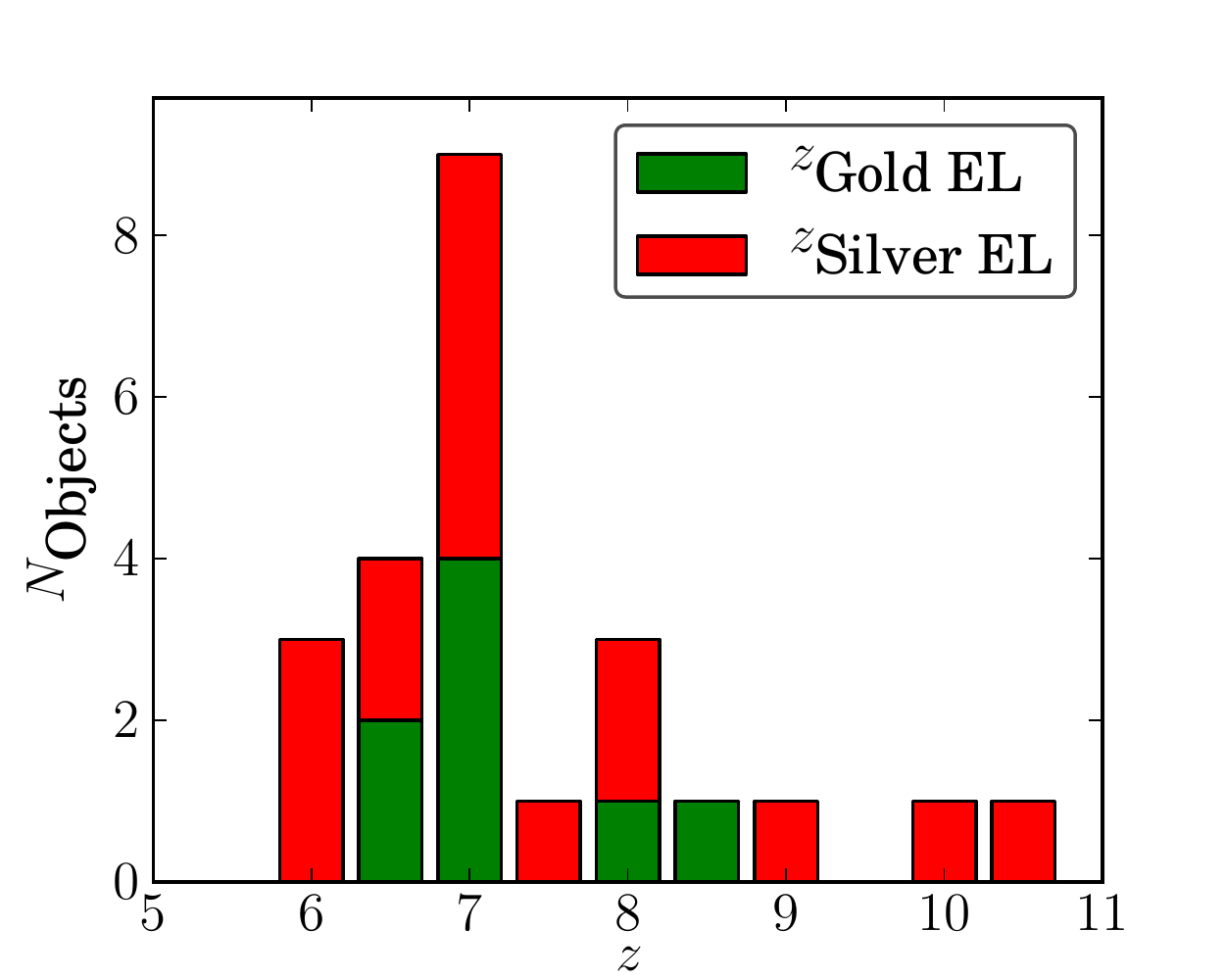}
\caption{The redshift distribution of the Gold (orange), Silver (gray), Silver\_EL (red) and Gold\_EL (green) samples from Tables~\ref{tab:dropouts} and \ref{tab:dropouts_EL}.
In both panels, the distributions are stacked to show the total number of sources in each bin.
For the Gold and Silver sample the redshift from the photometric selection(s) is used, whereas the redshifts for the EL samples correspond to the redshifts of the emission lines listed in Table~\ref{tab:dropouts_EL}, assuming that they are \lya.
}
\label{fig:zhist}
\end{center}
\end{figure*}

The Gold, Gold\_EL, Silver, and Silver\_EL objects are marked by the orange, green, gray, and red circles on each of the color composites shown in Figure~\ref{fig:clusters}.
The redshift distributions of the samples are shown in Figure~\ref{fig:zhist}. 
Here the mean redshift of the selection(s) is used for the Gold and Silver samples (Table~\ref{tab:dropouts}), whereas for the Gold\_EL and Silver\_EL samples (Table~\ref{tab:dropouts_EL}) we use the redshift corresponding to the emission line wavelengths listed in the `$\lambda_\textrm{lines} \pm50$\AA' column.

\section{Flux Limits and Equivalent Widths}\label{sec:flimandEW}
\begin{figure}
\begin{center}
\includegraphics[width=0.49\textwidth]{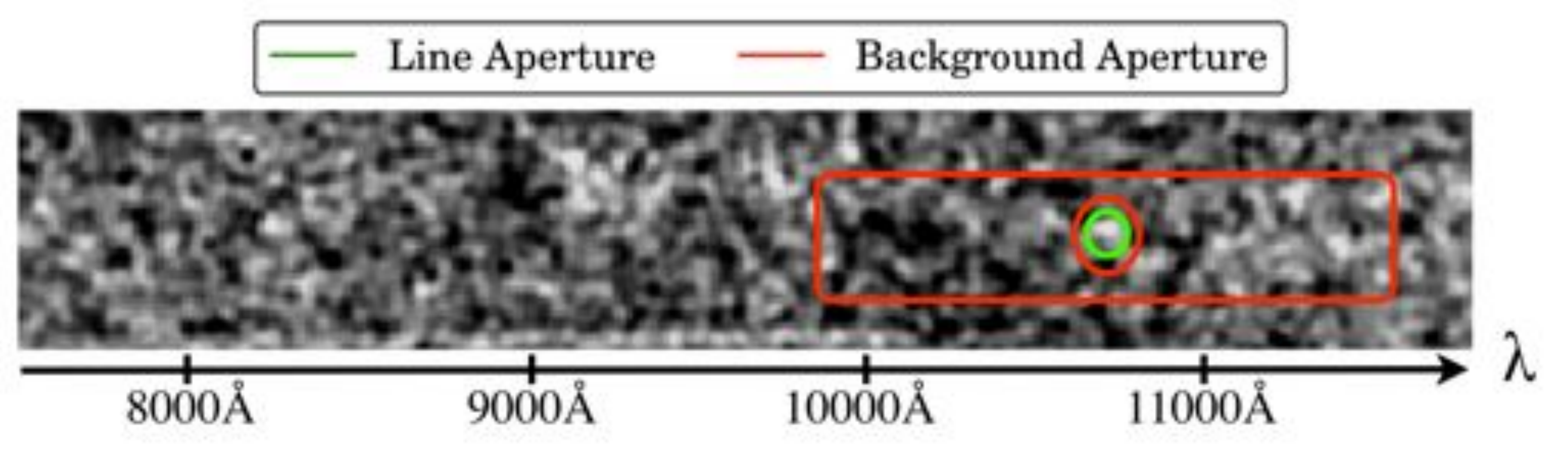}
\caption{Example of the emission line and background apertures used when estimating the emission line fluxes (and 1$\sigma$ flux sensitivities) of the GLASS spectra of \Rthirteen\_00627 shown in Figure~\ref{fig:fobjects}.
The line flux is obtained by accumulating (`integrating') the flux in the green ellipsoidal line aperture ($E_\textrm{line}$). 
The red background aperture (excluding the red ellipse around the line aperture) is used to normalize the background level of the contamination-subtracted spectrum to account for over/under subtraction of the local background (and contamination).}
\label{fig:fluxap}
\end{center}
\end{figure}

To quantify the emission line detections and non-detections, we estimate the line fluxes, emission line rest-frame equivalent widths, and the 1$\sigma$ line flux sensitivities. The rest-frame equivalent widths defined by
\begin{equation}\label{eqn:EW}
\textrm{EW} = \frac{ f_\textrm{line} } { f_\textrm{cont.}  \times (1+z)}    \; ,
\end{equation} 
were estimated based on the extracted two-dimensional spectra.
The `integrated' line flux, $f_\textrm{line}$, was estimated in two-dimensional ellipsoidal apertures adjusted for each individual object based on the extent of the line and the contamination (subtraction residuals) optimizing S/N and is given by 
\begin{equation}
f_\textrm{line} = \sum_i^{E_\textrm{line}} f_i - f_\textrm{bck}    \; ,
\end{equation}
where $E_\textrm{line}$ refers to the number of pixels in the ellipsoidal aperture used to enclose the line.
For the EL samples, $E_\textrm{line}$ has a median size of 66 pixels.
The line flux is corrected for background (and contamination) over/under subtraction, mainly due to the intra-cluster-light which varies strongly across the field-of-view, by adjusting the fluxes by the median background flux per pixel in a `background aperture' defined around the emission line for each spectrum, $f_\textrm{bck}$. 
An example of the line and background apertures used for \Rthirteen\_00627 is shown in Figure~\ref{fig:fluxap}.

In equation~(\ref{eqn:EW}) $f_\textrm{cont.}$ is the continuum level estimated from the ancillary broad band photometry given by
\begin{equation}
\frac{f_\textrm{cont.}}{ \left[10^{-17}  \textrm{erg/s/cm}^{2} \textrm{/\AA} \right] }  
= \frac{10^{-0.4 m_\textrm{AB}} \; \times \; 3 \times 10^{-1.44} }{ \left(\lambda_\textrm{obs} / [\textrm{\AA}] \right)^2}    \; ,
\end{equation}
with $m_\textrm{AB}$ being the F140W broadband magnitude.

We estimate 1$\sigma$ flux limits using the same approach, but replacing $f_\textrm{line}$ in equation~(\ref{eqn:EW}) with the uncertainty on the integrated flux given by
\begin{equation}
\sigma_\textrm{line} = \sqrt{\sum_i^{E_\textrm{line}}  \sigma_i^2}    \; .
\end{equation}

From the individual GLASS spectra we estimated the 1$\sigma$ flux limits for the Gold and Silver samples in Table~\ref{tab:dropouts}.
The 1$\sigma$ flux sensitivities were estimated using a spectral extraction aperture of roughly 5~(spatial) by 3~(spectral) native pixels which corresponds to $\sim$0.6\arcsec$\times$100\AA{} similar to what was used by \cite{Schmidt:2014p33661}, and were calculated at the wavelength of the mean redshift of the photometric selections given in the `$z_\textrm{sel.}$' column in Table~\ref{tab:dropouts}.
All spectra were subtracted a model of the contamination prior to estimating the flux limit and correcting the background offset.
In a few cases the spectra were hampered by severe contamination and the model subtraction was not ideal. 
These flux limits are potentially affected by the contamination level, despite our attempt to account for any offsets by adjusting the background of each individual spectrum.

\begin{figure}
\begin{center}
\includegraphics[width=0.49\textwidth]{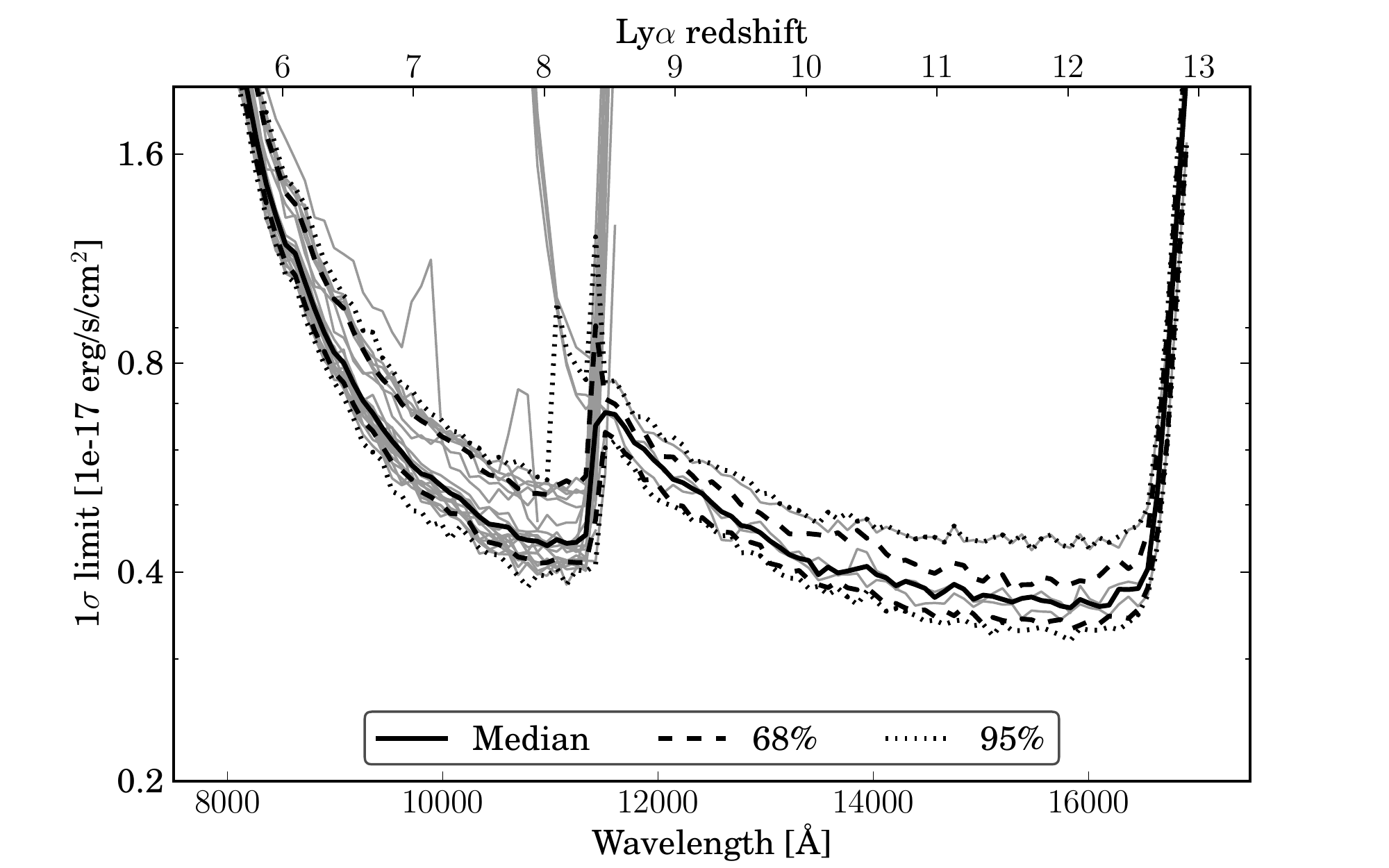}
\caption{The 1$\sigma$ sensitivity curves for the Gold sample in Table~\ref{tab:dropouts}. 
The dashed and dotted lines show the 68\% and 95\% spread of the limits for the individual spectra, whereas the thick solid line shows the median 1$\sigma$ sensitivity.
The limits have not been corrected for the lensing magnification of each object, $\mu$, listed in Table~\ref{tab:dropouts}.
Including the lensing magnification will improve the line flux sensitivity by a factor $\mu$. 
The gray curves correspond to the line flux sensitivity of 25 (to avoid over-crowding the plot) individual spectra from a single PA.
Combining the spectra from the two GLASS PAs for each object further decreases the noise level by a factor $\sqrt{2}$ at all wavelengths.
}
\label{fig:fluxlimits}
\end{center}
\end{figure}

By estimating the 1$\sigma$ flux limits stepping through the full wavelength range of the G102 and G141 grisms, we estimate the line flux sensitivity of the GLASS spectra as shown for the Gold sample in Figure~\ref{fig:fluxlimits}. 
These limits are in good agreement with the preliminary curves shown by \cite{Schmidt:2014p33661} and show that each spectrum reaches roughly $5\times10^{-18}$erg/s/cm$^2$ over the G102 and G141 wavelength range.
Combining the spectra of each object from the two individual GLASS PAs further strengthens this limit to $\sim3.5\times10^{-18}$erg/s/cm$^2$ (a factor of $\sqrt{2}$ better). 
These limits have not been corrected for lensing magnification, which will further improve the \emph{intrinsic} (as opposed to observed) flux sensitivity by a factor $\mu$. 
The lensing magnification for each object obtained from the HFF lensing web tool\footnote{\url{http://archive.stsci.edu/prepds/frontier/lensmodels/}} for the HFF clusters 
is tabulated in the $\mu$ columns of Table~\ref{tab:dropouts} and \ref{tab:dropouts_EL}.
Here we list the median lensing magnification from the available models. 
The quoted uncertainties correspond to the range of models, ignoring the maximum and minimum magnification, which essentially corresponds to the $68$\% ($\sim1\sigma$) range.
In this way, these errors minimize the effect of outliers and catastrophic error estimates in the lensing models.

As the \lya\ emission is expected to be more extended than the continuum flux
\citep[e.g.,][]{Steidel:2011p35184,Finkelstein:2011p36860,Laursen:2011p36006,Matsuda:2012p36912,Momose:2014p35332,Wisotzki:2015p41489}
 it is useful to also estimate the limiting flux for a larger aperture of, e.g., 
10~(spatial) by 6~(spectral) native pixels. 
In this case the 1$\sigma$ flux sensitivities of the GLASS spectra shown in Figure~\ref{fig:fluxlimits} essentially becomes shallower by a factor of roughly $\sqrt{60\textrm{pixels}/15\textrm{pixels}}=2$.

For the Gold\_EL and Silver\_EL samples we give the equivalent width
together with the measured line fluxes in Table~\ref{tab:dropouts_EL}.
These were estimated in rest-frame assuming that the detected features in the GLASS spectra are \lya. 
The assumed \lya\ redshift is given in parenthesis after the $z_\textrm{Sel.}$ redshift in Table~\ref{tab:dropouts_EL}
and the distribution is shown in the right panel of Figure~\ref{fig:zhist}.

\section{Automated Line Detection In the GLASS Spectra}\label{sec:pline}

To complement the visual inspection described above, we performed an
automated line search, utilizing the newly developed Bayesian
statistical line detection framework described by
\cite{Maseda:2015p39932}.  The fundamental assumption of the framework
is that the morphology of the emission line follows the NIR
morphology of the object determined by the direct images measured in overlapping filters.  
The likelihood of the observed two-dimensional spectra given a line flux, is then
estimated based on a noise model. Assuming a uniform prior for the
fraction of continuum flux in the line $A = f_\textrm{emission line} /
f_\textrm{rest-frame UV image}$, this yields the posterior
distribution function for the presence of a line at any given
wavelength. The fraction $A$ is allowed to be negative, so the probability 
$ p(A>0) = \int_{0}^\infty p(A) dA/\int_{-\infty}^\infty p(A) dA $
calculated at $\lambda$ gives the probability of the existence of an emission
at that wavelength.  For more details on the Bayesian line detection
software we refer the reader to \cite{Maseda:2015p39932}.

We applied the \cite{Maseda:2015p39932} framework to the Silver, Gold,
Gold\_EL, and Silver\_EL samples.  For each object, this resulted in
four (two grisms $\times$ two PAs) probability curves for the
detection of lines at each wavelength.  We combined these four curves
to a single probability profile by calculating
\begin{equation}
p_\textrm{comb.}(A>0) = \frac{\int_{0}^\infty \prod_i p(A_i)
\;dA}{\int_{-\infty}^\infty \prod_i p(A_i) \;dA}    \; ,
\end{equation}
where the product is over the $i$ spectra at the given wavelength.  By
allowing for small shifts ($<\pm25$\AA) in wavelength of each curve,
maximizing the 2$\sigma$ peaks of $p_\textrm{comb.}(A>0)$, we account
for any uncertainties in the GLASS reduction wavelength solutions.

All individual and combined $p$-curves were searched for
high-significance peaks and visually inspected at $\lambda =
8500-16500$\AA.  Spurious line detections from contamination
subtraction residuals and at the low-sensitivity edges of the spectra
were discarded.  In Table~\ref{tab:plinevals} of
Appendix~\ref{sec:plinevals} we have listed the maximum probabilities
around the visually identified lines from Table~\ref{tab:dropouts_EL},

\section{Statistical analysis of the Ly$\alpha$ detections}\label{sec:stats}

In this section we aim to assess the statistical properties of the
sample of \lya\ detections in comparison with those found by other
studies and the numbers predicted by theoretical models. In order to
carry out this comparison, we first estimate the completeness
and purity of 
our Lyman break galaxy samples with \lya\ detections, 
as described in
Section~\ref{ssec:comp}. Then, in Section~\ref{ssec:stats}, we present the
comparison.
We note that the varying depth of the ancillary data and the different photometric pre-selections used to assembled the Gold and Silver samples, do not affect the statistical analysis of the line emitters presented in this section.
As long as we have a homogeneous limit in flux for the emission lines, the statistics are unaffected by the pre-selections.

\subsection{Completeness and Purity of Visually Selected Emission Line Samples}\label{ssec:comp}

The automated procedure described in the previous section allows us to
estimate how many high-significance emission lines were missed by the
visual inspection. Of course this estimate of completeness only
applies to the line emission with morphology well described by that of
the continuum. Lines that are significantly more extended, compact, or
offset with respect to the continuum might have lower significance and
thus be missed by both the automated and visual procedure or
identified only by the visual procedure.  As we will discuss further
in Section~\ref{sec:individualobj}, we are certain that some emission
lines have not been picked up by our conservative visual inspection,
since at least one Gold object, RXJ1347\_01037, has been confirmed to be
a \lya\ emitter by follow-up spectroscopy, while it had only been
identified by one visual inspector and is therefore not included in
the Gold\_EL sample.  Conversely, we expect the visual inspection
procedure to pick up \lya\ offset from the continuum emission which
(by definition) will be missed by the automatic line detection.  Such
offsets are seen at both lower \citep{Shibuya:2014p40120} and higher
redshift \citep{Jiang:2013p40145} and are therefore expected at these
redshifts, and indeed we may have observed this in the case of
MACS2129\_00899 which is also discussed in
Section~\ref{sec:individualobj}.
The feature in MACS2129\_00899 has a low $p$-value
(cf. Table~\ref{tab:plinevals}), even though it is clearly detected in
the G102 grism (see Figure~\ref{fig:z8obj}).  However, as it is
somewhat offset from the continuum and its morphology is different,
we do not expect to pick it up with the automated detection software.

We can also estimate the purity, i.e. one minus the fraction of
contaminants \footnote{Here, contaminants refers to both faint low-$z$ line emitters 
resembling \lya\ emitters at high redshift, as well as features 
in the spectra not necessarily stemming from true astronomical sources.
We note that all obvious defects and 0th order images were removed 
from the contaminants before purity and completeness was estimated. 
The fraction of faint low-$z$ line emitters is confirmed to be small from 
a stack of the emission line sources as described in Section~\ref{sec:stack}.},
of the visual emission line sample by using the same
automated detection software \citep{Maseda:2015p39932}. By running the
code on parts of sky where there are no photometrically detected
dropouts we estimate how many contaminants to expect, including both
noise spikes and pure line emitters with no continuum. In practice, in
order to mimic the data quality as closely as possible, for each
dropout we ran the line detection software on a trace offset 10 pixels
above and 10 pixels below the main target, along the spatial direction. We 
counted the occurrence of 3$\sigma$ detections.  We find
spurious detections above 3$\sigma$ in 4/26 of the offset traces in the spectra
of the 26 objects suitable for this test (spectra clean from contamination
subtraction residual and defects in the center as well as in the
offset traces). Conservatively, we assume that those are true false
positives, even though some might be true emissions lines, associated
with objects that are too faint in the continuum to be detected in our
images.

By carrying out the calculations described in detail in
Appendix~\ref{app:purcomp}, based on the output of the automated
detection software, we can estimate our visual completeness and
purity. For the Gold sample, the 3$\sigma$ completeness (i.e. how many
of the actual line emitters with flux above the threshold, we identify)
is in the range 40-100\%, while the purity is in the range 60-90\%
(i.e. 10-40\% of the line detections are spurious). For the Silver 
sample the completeness and purity are the same within the
uncertainties (40-100\% and 65-90\%, respectively). Deeper
spectroscopic follow up is needed to improve these estimates.

\subsection{Statistics of the \lya\ Emitters}
\label{ssec:stats}

Armed with estimates of the completeness and purity derived in the
previous section we can proceed with a statistical comparison of our
sample to the expectations based on previous work. Before carrying out
the comparison, we emphasize that the sample size of line emitters is
relatively small and the completeness and purity estimates are
uncertain, and therefore no strong conclusions can be drawn at this
stage. 
Furthermore, given the heterogeneity of the photometric
selection it is premature to carry out a detailed inference of the
\lya\ optical depth based on the individual properties of each object,
as described by \cite{Treu:2012p12658}. We thus leave a detailed
analysis for future work, when the full GLASS dataset combined with
the full depth HFF images (and Spitzer IRAC photometry) have been 
analyzed to allow for a homogenous
photometric pre-selection.

The model presented by \cite{Treu:2015p36793} allows us to estimate
how many \lya\ emitters we would have expected to detect in the six
GLASS clusters, given the detection limits presented in this paper.
Briefly the model adopts the \cite{Mason:2015p40685} luminosity
function for the UV continuum, associates \lya\ to the UV magnitude
following the conditional probability distribution function inferred
by \cite{Treu:2013p32132} and \cite{Pentericci:2014p34725} at
$z\sim7$, and then accounts for the effects of cluster magnification
by randomly generating sources in the source plane and lensing them
through actual magnification maps. Based on the model, we expect to
detect 2-3 \lya\ lines per cluster at $z\gtrsim7$ with flux above
10$^{-17}$erg/s/cm$^{2}$ (our 3$\sigma$ limit for galaxies imaged at
two position angles).  Thus for six clusters we would have expected
roughly 12-18 3$\sigma$ detections. The Gold\_EL sample
consists of 8 detections and the Silver\_EL sample contains 16
additional detections.  Formally 3 and 9 of these emission lines are
3$\sigma$ detections cf. Table~\ref{tab:dropouts_EL}).  In
Table~\ref{tab:ELstat} we summarize the expected sample sizes of
true \lya\ emitters applying the estimated completeness and purity
corrections described above with and without poisson statistics on the
samples \citep[cf.][]{Gehrels:1986p40183}.  We consider both a
pessimistic scenario using the lower bounds of both the completeness
and purity ranges (``Low C \& P'' column), and a more optimistic
scenario using a completeness of 100\% and a 90\% purity of our
samples (``High C \& P'' column).

From Table~\ref{tab:ELstat} it is clear that a quantitative
comparison is very difficult to make, and depends strongly on the
level of contamination and purity of the samples. 
However, in summary the formal 3$\sigma$ Gold\_EL sample is below the expected number of line emitters.
The combined 3$\sigma$ sample
Silver\_EL+Gold\_EL agrees with the expected number of line emission
within the 68\% and 95\% confidence levels, as does the majority of
the 3$\sigma$ Silver\_EL sampled and the individual samples 
when all line detections, irrespective of S/N, are included. 
It is encouraging that the numbers are in rough agreement
with the model calibrated on previous measurements, indicating that we
are not grossly over or under-estimating the number of contaminants and
the incompleteness. Better defined photometric selections and more
spectroscopic follow-up is needed before any firm conclusions can be
drawn. In a year or two, with better data in hand, it will be possible
to carry out a detailed statistical analysis as outlined by
\cite{Treu:2012p12658}, and reach quantitative conclusions.

\tabletypesize{\scriptsize} \tabcolsep=0.15cm
\begin{deluxetable*}{l|ccc|ccc} \tablecolumns{7}
\tablewidth{0pt}
\tablecaption{Emission Line Number Statistics}
\tablehead{
{ } & \multicolumn{3}{c|}{All} & \multicolumn{3}{c}{Formal 3$\sigma$ detections cf Table~\ref{tab:dropouts_EL} } \\ 
Sample & Detections & Low C \& P & High C \& P & Detections & Low C \& P & High C \& P
}
Gold\_EL 				&	8.0		&	12.0		&	7.2		&	3.0		&	4.5		&	2.7		\\
Silver\_EL 			&	16.0		&	26.0		&	14.4		&	9.0		&	14.6		&	8.1		\\
Silver\_EL+Gold\_EL		&	24.0		&	28.0		&	21.6		&	12.0		&	18.9		&	10.8		\\
\hline
\multicolumn{7}{c}{} \\
\multicolumn{7}{c}{Poisson Statistics Ranges: 68\% [95\%] Confidence Levels \citep[cf.][]{Gehrels:1986p40183} } \\
\hline
{ } & \multicolumn{3}{c|}{All} & \multicolumn{3}{c}{Formal 3$\sigma$ detections} \\
Sample & Detections & Low C \& P & High C \& P & Detections & Low C \& P & High C \& P \\
\hline
Gold\_EL 				&	5--12 [3--16]		&	9--17 [6--21]		&	4--11 [3--14]		&	1--6 [1--9]			&	3--8 [2--12]		&	1--6 [1--9]]			\\
Silver\_EL 			&	12--21 [9--26]		&	21--32 [17--38]		&	10--19 [8--23]		&	6--13 [4--17]		&	11-20 [8--25]		&	5--12 [3--16]		\\
Silver\_EL+Gold\_EL		&	19--30 [15--36]		&	23--35 [19--40]		&	17--28 [14--33]		&	9--17 [6--21]		&	15--24 [12--30]		&	8--15 [5--20]		
\enddata
\tablecomments{
The top cells list numbers not accounted for poisson statistics. The bottom cells list the corresponding ranges including poisson noise.
The `detections' column refers to the number of potential \lya\ emitters listed in Table~\ref{tab:dropouts_EL}. 
`Low C \& P' refers to an assumed low contamination and low purity of 40\% (40\%) and 60\% (65\%) for the Gold\_EL (Silver\_EL) sample, respectively.
`High C \& P' refers to an assumed high contamination and high purity of 100\% and 90\%, respectively, for both the Gold\_EL and Silver\_EL sample. 
}
\label{tab:ELstat}
\end{deluxetable*}

We conclude that our results are consistent with the predictions of
simple empirical models based on previous measurements of the \lya\
emission probability at $z\sim7$. Therefore, our findings are
consistent with previous work, that shows that the probability of
\lya\ emission is lower at $z\gtrsim7$ than at $z\sim6$
\citep[e.g.,][]{Pentericci:2011p27723,Pentericci:2014p34725,Schenker:2012p34406,Schenker:2014p35145,Treu:2013p32132,Tilvi:2014p35476,Caruana:2014p32713}.
In the future, larger samples, deep spectroscopic follow-up and a
homogenous photometric pre-selection will allow us to reduce the
uncertainties and hopefully separate the sample in $z\sim7$ and
$z\sim8$ candidates.

\section{Note on four Individual objects}
\label{sec:individualobj}

In the following we describe RXJ1347\_01037 which has been
independently confirmed to be a galaxy at $z=6.76$ with Keck DEIMOS
spectroscopy, MACS2129\_00899, a very promising $z=8.1$ candidate, 
and the two potential $z\sim10$ objects from table~\ref{tab:dropouts_EL} MACS1423\_01018 and RXJ2248\_00404 which are possible low-redshift contaminants.

\subsection{RXJ1347\_01037}\label{sec:rxj1347_1037}

RXJ1347\_01037 has been independently confirmed to be a line emitter
from Keck-DEIMOS observations as presented by
\cite{Huang:2015p39241}. They detect emission at $\sim9440$\AA.

\begin{figure}
\begin{center}
\includegraphics[width=0.49\textwidth]{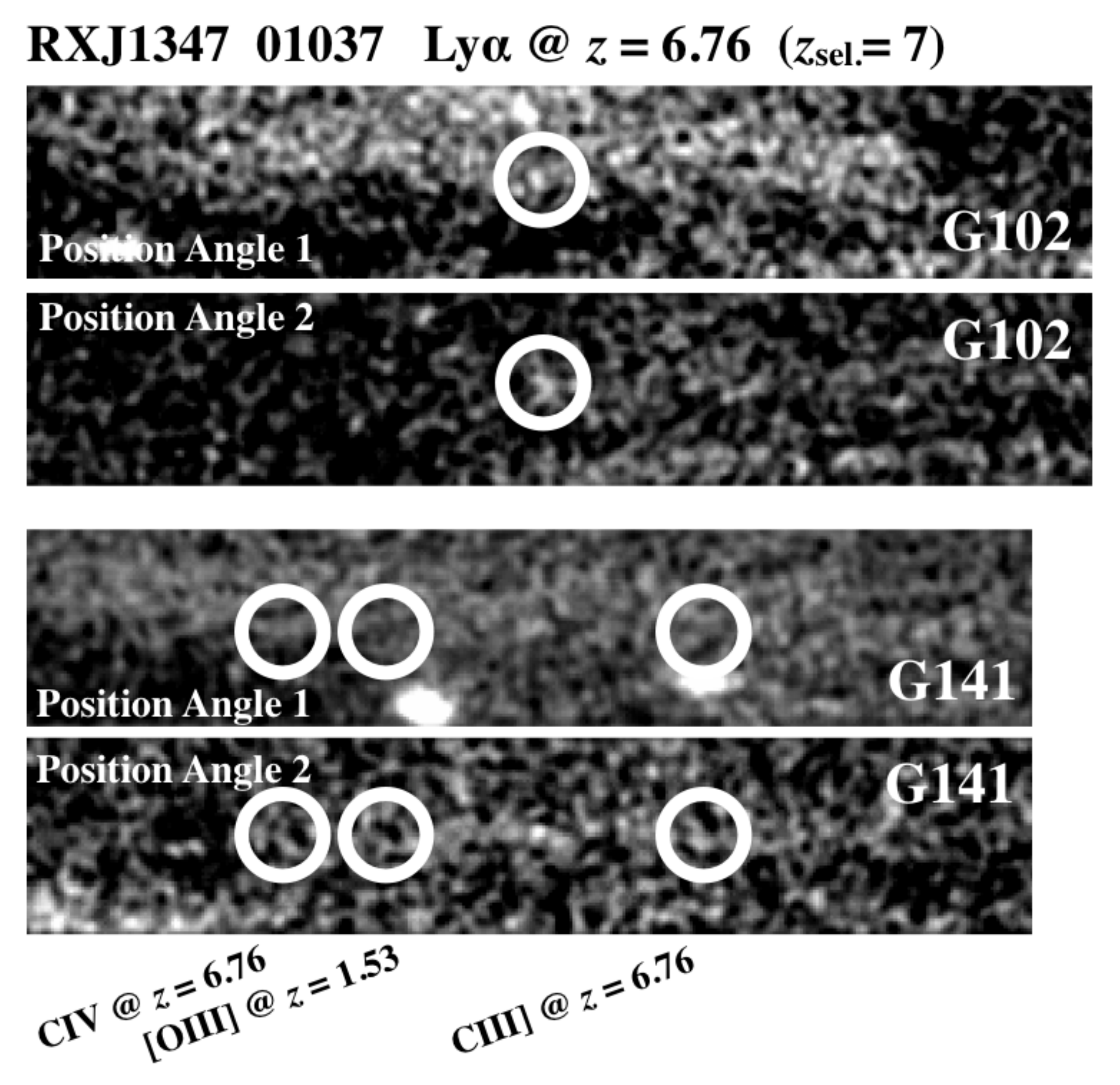}
\caption{The GLASS spectra of the confirmed \lya\ emitter at $z=6.76$ RXJ1347\_01037 
described in Section~\ref{sec:rxj1347_1037} and presented by \cite{Huang:2015p39241}.
The white circles in the G102 spectra mark the position of \lya.  The
central white circles in the G141 spectra mark the position of
[OIII]$\lambda$5007, if the line was [OII]$\lambda$3727 at $z=1.35$.  
The lack of [OIII] emission supports the interpretation that the G102
emission line is \lya, given that a low-metallicity object at $z=1.35$
would have [OIII]/[OII]$ > 1$.  
The left and right-most
white circles in the G141 spectra marks the location of
CIV$\lambda$1549\AA{} and CIII]$\lambda$1909\AA{} at $z=6.76$, respectively.  
We do not detect any significant CIV and CIII] emission from this object
with line ratio limits of $f_\textrm{CIV} / f_\textrm{\lya}
\lesssim 0.36$ and $f_\textrm{CIII]} / f_\textrm{\lya}\lesssim 0.25$.
}
\label{fig:z7obj}
\end{center}
\end{figure}

RXJ1347\_01037 is in our photometric Gold sample. It does not appear
in the Gold\_EL, as the line was marked by only one inspector (at $\sim
9440$\AA\ in the G102 grism at both PAs). Figure~\ref{fig:z7obj} shows
all four GLASS spectra of RXJ1347\_01037 with the emission line marked in
the G102 spectra.

In the Keck-DEIMOS spectrum the blue side of the line falls on a sky
line residual.  Hence, even though the DEIMOS spectrum would resolve
the [OII] doublet at $z=1.53$, the identification of the feature as
\lya\ at $z=6.76$ from Keck is not fully conclusive 
(even though the asymmetric line profile is consistent with
it). Unfortunately, in the GLASS spectra the resolution is too low to
resolve the doublet. 

However, GLASS can confirm the line identification as \lya\ by virtue
of its NIR spectral coverage. If the detected line were
[OII]$\lambda$3727, we would expect to see [OIII]$\lambda$5007
emission, based on typical line ratios.  The DEIMOS wavelength
coverage is not sufficient to look for potential [OIII] emission.
GLASS in contrast has sufficient wavelength coverage in the G141 grism
(these spectra are also shown in Figure~\ref{fig:z7obj}).  We do not
detect any flux in the G141 spectra at $\lambda\sim12685$\AA\ (marked
by the central white circles in the G141 spectra in
Figure~\ref{fig:z7obj}) which would be the expected position of
[OIII]$\lambda$5007 at $z=1.53$.  If the object is a low metallicity
object at $z=1.53$ we expect [OIII]/[OII] $> 1$
\citep[e.g.,][]{Nagao:2006p38359,Maiolino:2008p35580} which is
certainly not the case.  
A high-metallicity galaxy would show high ratios of [OII]/[OIII] consistent 
with what is observed. However, as star forming galaxies will always have 
either [OIII] or H$\beta$ flux $>0.3\times$ the [OII] flux \citep{Jones:2015p40575},
the non-detection of H$\beta$ makes such a scenario very unlikely.
Combining the fluxes in the individual
spectra (see below) and using the 2$\sigma$ flux limits at the
location of [OIII], the limit on the [OIII]/[OII] ratio from the GLASS
spectra becomes $f_{2\sigma\textrm{lim., [OIII]}} / f_\textrm{[OII]}
\lesssim 0.32$.  
Furthermore, the automatic line detection mentioned in Section~\ref{sec:pline} 
assigns a combined probability $p(A>0) =  0.999939$, 
which corresponds to a 4.01$\sigma$ detection, of a line at $9440\pm50$\AA.
Based on this probability, the non-detection of [OIII] and the [OIII]/[OII] flux ratio limit
we conclude that the line detected in the GLASS and
Keck-DEIMOS spectra is \lya\ at $z=6.76$, in agreement with the
conclusion based on the line profile by \cite{Huang:2015p39241}, and
strongly favored by the photometry.  Given the \lya\ emission and the
redshift, the GLASS wavelength coverage allows us to search for
CIV$\lambda$1549\AA{} and CIII]$\lambda$1909\AA{} emission at
12020\AA\ and 14815\AA\ (marked by the left- and right-most white
circles in the G141 spectra in Figure~\ref{fig:z7obj}).  We do not
detect any significant CIV or CIII] emission from RXJ1347\_01037 in
the GLASS spectra.

Estimating the \lya\ line flux and equivalent width from the GLASS G102
spectra, we find (fluxes not corrected for magnification)
\begin{eqnarray}
f_\textrm{line} 			&=& 2.1\pm0.8  \times 10^{-17}\textrm{erg/s/cm}^2 \\
\textrm{EW}_\textrm{\lya} &=& 61\pm24\textrm{\AA} 
\end{eqnarray}
for the S/N=2.6 line (PA 1 in Figure~\ref{fig:z7obj}), and 
\begin{eqnarray}
f_\textrm{line} 			&=& 3.1\pm0.7  \times 10^{-17}\textrm{erg/s/cm}^2 \\
\textrm{EW}_\textrm{\lya} 	&=& 88\pm22\textrm{\AA} 
\end{eqnarray}
for the S/N=4.1 line (PA 2 in Figure~\ref{fig:z7obj}), 
resulting in a combined line flux and equivalent width of
\begin{eqnarray}
f_\textrm{line} 			&=& 2.6\pm0.5  \times 10^{-17}\textrm{erg/s/cm}^2 \\
\textrm{EW}_\textrm{\lya} 	&=& 74\pm16\textrm{\AA}    \; .
\end{eqnarray}

These two estimates are in mutual agreement, but in tension with the
line flux and equivalent width estimated from the DEIMOS spectrum by
\cite{Huang:2015p39241}.  They find a line flux of $f_\textrm{line} =
7.8\pm0.7 \times 10^{-18} \textrm{erg/s/cm}^2$ and a \lya\ 
equivalent width of $\textrm{EW}_\textrm{\lya} = 26\pm4$\AA{}.
The GLASS and DEIMOS fluxes taken at face value differ by 4$\sigma$.
We expect this difference to be caused by systematic uncertainties in
the ground based DEIMOS spectrum, including those stemming from
a combination of slit losses, absolute spectrophotometric calibration and 
sky line subtraction.

From the \lya\ estimates we find the following upper limits on the rest-frame UV emission line ratios from RXJ1347\_01037:
\begin{eqnarray}
f_{2\sigma\textrm{lim., CIV}} / f_\textrm{\lya} 		&\lesssim& 0.36 \\
f_{2\sigma\textrm{lim., CIII]}} / f_\textrm{\lya}   		&\lesssim&  0.25    \; .
\end{eqnarray}
Here we have again used the 2$\sigma$ flux limit on CIV and CIII].

\subsection{MACS2129\_00899}\label{sec:macs2129_899}

\begin{figure}
\begin{center}
\includegraphics[width=0.49\textwidth]{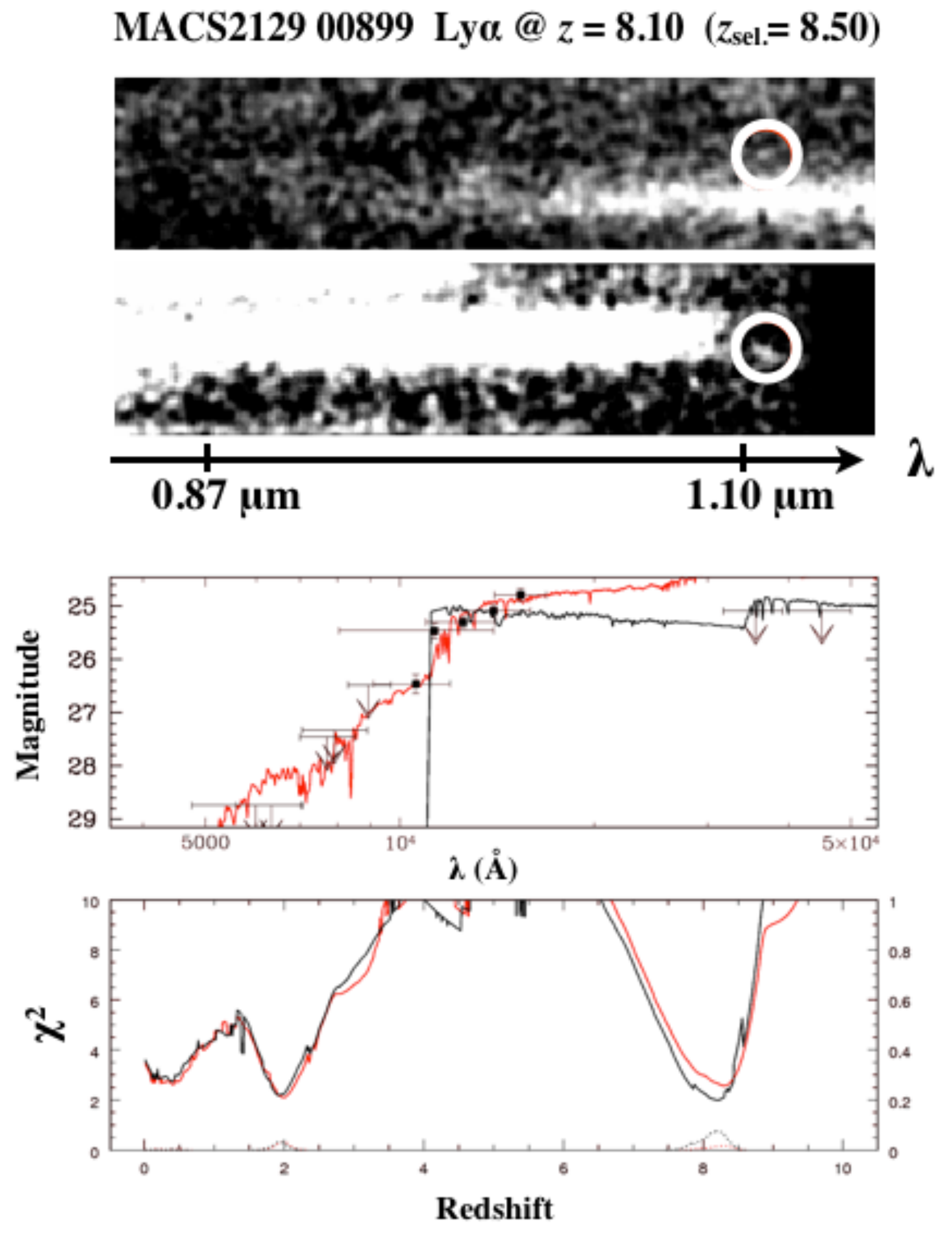}
\caption{The GLASS G102 spectra (top panels), photometry (central panel) and photometric redshift estimate $\chi^2$ curve (bottom panel) for the candidate \lya\ emitter at $z=8.1$ MACS2129\_00899.
The photometry includes CH1 and CH2 IRAC 1$\sigma$ upper limits from
SURFS-UP
\citep{Bradac:2014p34085} obtained following \cite{Huang:2015p39241}.
The black $\chi^2$ curve in the bottom panel includes F160W, whereas the red curve does does not (F160W has potential contamination and therefore uncertain photometry). 
In both cases there are valid photometric redshifts around $z\sim8$ (black spectral energy distribution over-plotted the photometry in center panel) and $z\sim2$ (red spectral energy distribution over-plotted the photometry in center panel) with marginal statistical difference. 
The $z\sim2$ solution over-estimates the 1$\sigma$ IRAC constraints.
If the line was [OII]$\lambda$3727 at $z=1.97$ we would expect to see [OIII] at roughly 14870\AA. 
We do not detect any [OIII] emission in the GLASS G141 spectra, consistent with the $z\sim8$ solution.
}
\label{fig:z8obj}
\end{center}
\end{figure}

Another object worth high-lighting is MACS2129\_00899.  It is a high
confidence \lya\ emitter candidate at $z=8.10$ which shows emission
lines in both of the G102 spectra at 11065\AA.  This wavelength is
also covered by the G141 grisms, albeit at low sensitivity (see
Figure~\ref{fig:fluxlimits}).  Despite the low sensitivity and
relatively high contamination there appears to be a marginal detection
of the line in one of the G141 spectra as well.  The estimated flux
and equivalent widths of the G102 lines are quoted in
Table~\ref{tab:dropouts_EL} and are in mutual agreement with combined
line flux and \lya\ equivalent width of
\begin{eqnarray}
f_\textrm{line} &=& 1.0\pm0.3 \times 10^{-17}\textrm{erg/s/cm}^2 \\
\textrm{EW}_\textrm{\lya} 	&=& 59\pm21\textrm{\AA}     \; .
\end{eqnarray}
In the top panels of Figure~\ref{fig:z8obj} we show the G102 spectra
with the \lya\ line marked by the white circles.  Consistent with the
photometric selection criteria listed in Table~\ref{tab:dropouts_EL}, the
photometric redshift posterior distribution function (in the form of $\chi^2$)
shown in Figure~\ref{fig:z8obj}  peaks at $z\sim8$. 
The spectral energy distribution templates fitting the photometry best are 
also shown in Figure~\ref{fig:z8obj}.
This redshift estimate comes from an independent fit using a current 
version of \verb+zphot+ \citep{Giallongo:1998p40150} based on independent 
HST and Spitzer photometry from SURFS-UP \citep{Bradac:2014p34085} 
obtained following \cite{Huang:2015p39241}.
As is often the case for $z\sim8$ galaxy candidates, a local
$\chi^2$-minimum is also seen at redshift $\sim$2.  In this case, the
emission line could be [OII]$\lambda$3727 at $z=1.97$, and the
observed break would be the 4000\AA\ break instead of the \lya\ break.
However, if the lower redshift solution were correct,
[OIII]$\lambda$5007 would fall at 14870\AA.  As was the case for
RXJ1347\_01037, we do not detect any [OIII] emission in the GLASS G141
spectra, which supports the interpretation of the G102 emission
feature as \lya\ at $z=8.1$.  We do not detect any CIV at 14095\AA\ in
the G141 spectra for this sources either (CIII] will fall at 17371\AA\
which is outside the G141 wavelength coverage).  The limit on the
CIV/\lya\ flux ratio obtained from the GLASS spectra is
$f_{2\sigma\textrm{lim., CIV}} / f_\textrm{\lya} \lesssim 0.64 $,
where we have again used the 2$\sigma$ limiting flux for CIV.  This limit is
in good agreement with current estimates of CIV/\lya\ flux ratios at
intermediate and high redshift
\citep{Shapley:2003p36937,Erb:2010p36933,Stark:2014p36777,Stark:2015p39317,Stark:2015p36807},
which are generally less than 0.6.

If confirmed, this would be one of the highest redshift sources known to
date, together with the recent $z=8.7$ galaxy confirmed by \cite{Zitrin:2015p40322} and the $z=8.2$ $\gamma$-ray burst presented by \cite{Salvaterra:2009p41930} and \cite{Tanvir:2009p34056}.
However, due to the relatively low resolution of the \emph{HST}
grisms and the low S/N of the lines presented here, deep high
resolution spectroscopic follow-up is needed to confirm the
high-redshift nature of this source, or deeper photometry to further
improve the photometric redshift estimate.

\subsection{Two Potential $z\sim10$ Objects}\label{sec:z10}
\begin{figure}
\begin{center}
\includegraphics[width=0.49\textwidth]{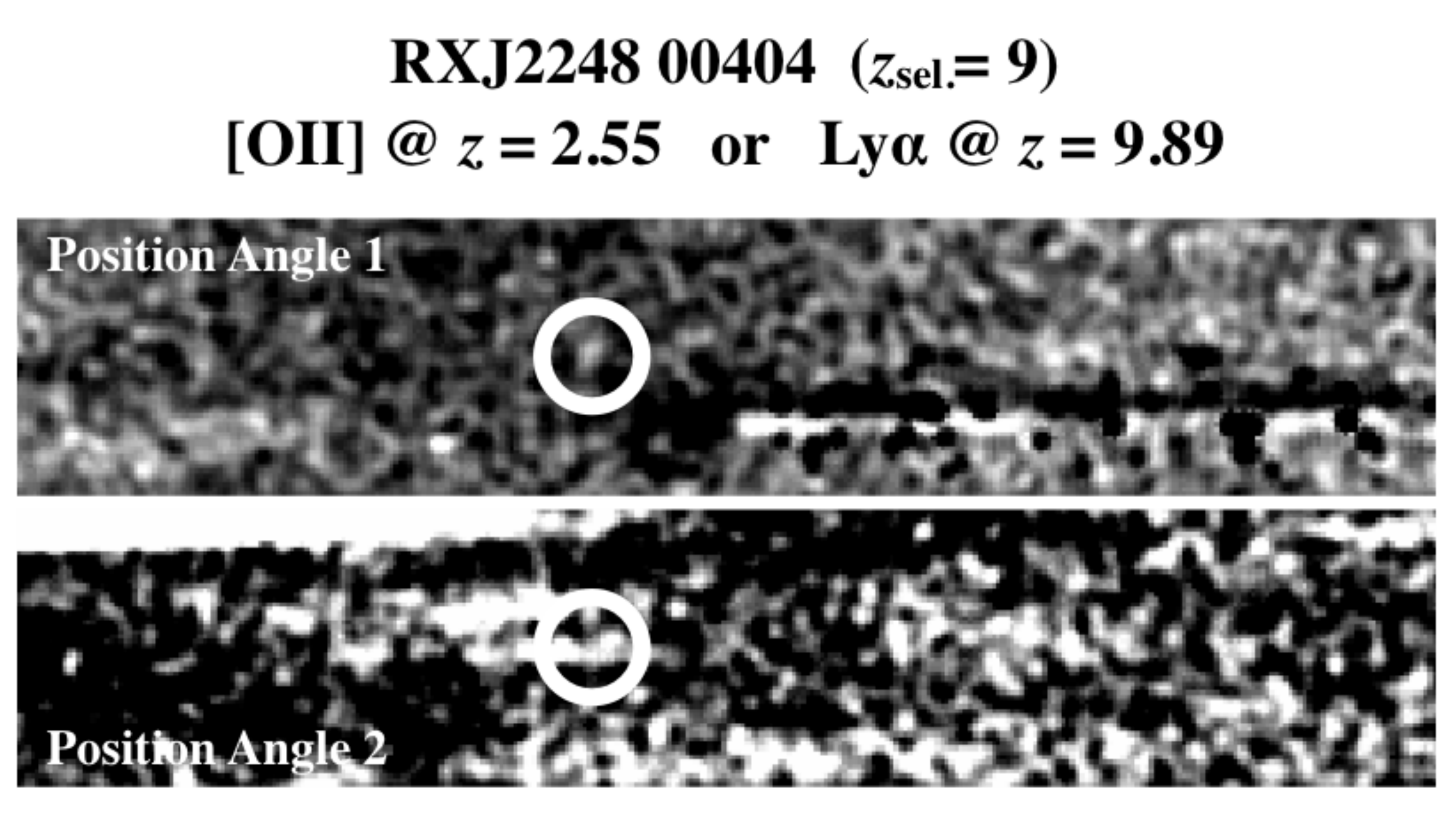}\\
\includegraphics[width=0.49\textwidth]{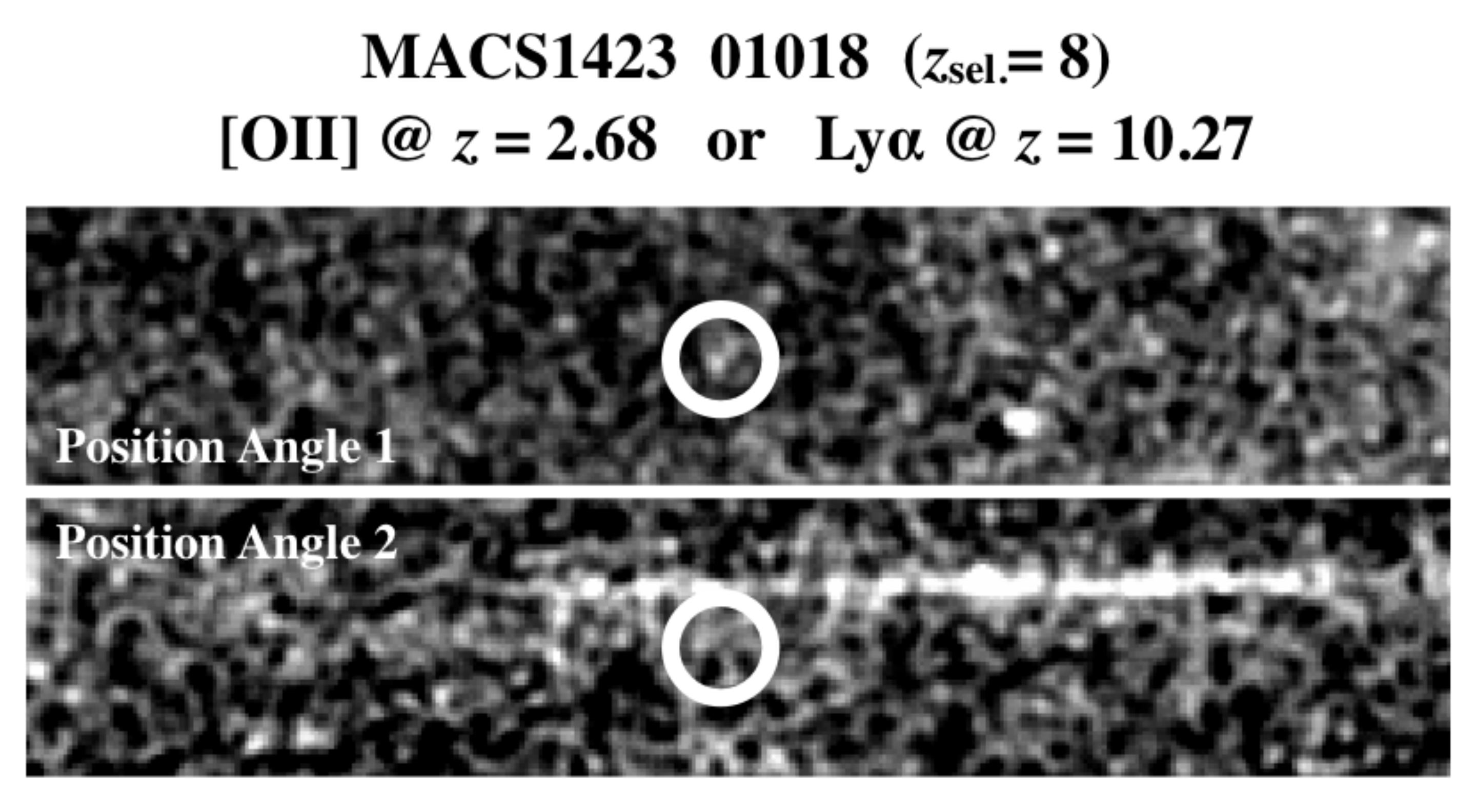}
\caption{
The GLASS G141 spectra of the two potential line emitters at $z\sim10$ from the Silver\_EL sample.
The white circles mark the detected line emission.
The objects were selected as Lyman break dropouts. 
However, highly likely low-redshift solutions in the EA$z$Y p($z$) suggest that the lines could also be [OII] at $z\sim2.6$.
Given the location of the lines at 1.32$\mu$m and 1.37$\mu$m the G141 spectrum does not cover [OIII], and the resolution does not allow us to resolve the potential [OII] doublet or detect the asymmetry of \lya.
With the current data we can therefore not rule out the high-redshift scenario.
}
\label{fig:z10obj}
\end{center}
\end{figure}

As presented in Table~\ref{tab:dropouts_EL}, RXJ2248\_00404 and MACS1423\_01018 of the Silver\_EL sample appear to have emission lines at 1.32$\mu$m and 1.37$\mu$m, respectively. 
If these lines are confirmed to be \lya\ this would place these objects at $z\sim10$.
Figure~\ref{fig:z10obj} shows the G141 spectra, marking the detected emission lines with white circles.
Both objects are selected as photometric dropouts, i.e. selected based on a few detections red-wards of the Lyman break and non-detections in bands blue-wards of the break.
In both cases the EA$z$Y photometric redshift distributions have highly probable solutions at $z\sim2$--3, and the photometry is therefore inconclusive as to whether the objects are at high redshift or low redshift.
If the emission lines are [OII] at $z=2.55$ and $z=2.68$ for RXJ2248\_00404 and MACS1423\_01018, respectively, this would agree with the EA$z$Y $p(z)$, and rule out the color selections placing them at redshift 8 and 9. 
In case the sources are at redshift 2--3, the drop in the NIR colors used to select them as high redshift galaxies could be attributed to the 4000\AA\ break as opposed to the Lyman break, which is known to be one of the main contaminants of Lyman break galaxy samples.
The resolution of the G141 grism is too low to resolve the [OII] doublet or detect the asymmetry of the \lya\ line, and we can therefore not distinguish between the two lines with the GLASS data. 
We do also not have the wavelength coverage to look for the [OIII] doublet, which would fall at 17775\AA\ and 18426\AA, respectively.
In summary, the two potential $z\sim10$ objects might be contaminats showing [OII] emission at $z\sim2.6$, but without follow-up spectroscopy or deeper imaging we cannot rule out the high-redshift \lya\ scenario.

\section{Physical Properties from \lya\ Emitters}\label{sec:ELimplication}

As described in the introduction, spectroscopically confirmed \lya\ emitters at high redshift (and the non-confirmations from follow-up campaigns) have proven very valuable for studying the early universe and the environment at the epoch of reionization \citep{Pentericci:2011p27723,Pentericci:2014p34725,Caruana:2012p27502,Caruana:2014p32713,Treu:2012p12658,Treu:2013p32132,Tilvi:2014p35476,Faisst:2014p34184}.
In particular, confirmed \lya\ emitters fix the redshift of the object resulting in improved prediction power from fitting stellar populations synthesis models to the photometry. 
Assuming a set of stellar population models \citep[e.g.,][]{Bruzual:2003p41996,Maraston:2005p41967} to generate spectral energy distributions for galaxies at the emission line redshift, and fitting them to the available photometry, can give estimates of physical quantities of the galaxies like, total stellar mass (the normalization between the observed flux and best-fit model), the star formation rate, metallicity and the age of the stellar populations, i.e. the galaxy 
\citep[e.g,][]{Labbe:2006p41953,Vanzella:2011p29486,Finkelstein:2013p32467,Coe:2013p26313,Coe:2014p35275,Huang:2015p39241,Zitrin:2015p40322,Oesch:2015p38504}.
When performing the spectral energy distribution fitting, assuming a dust law can furthermore predict the dust content of the galaxy. This can be directly compared to the measured UV spectral slope, if available from the data \citep[e.g.,][]{Oesch:2015p38504,Finkelstein:2015p37430,Bouwens:2015p34683}. 
Another direct comparison can be obtained from independently determining the star formation rate from scaling relations with the UV photometry \citep{Kennicutt:1998p9250,Madau:1998p41934}.
As part of our study of IRAC-detected high-redshift galaxies presented by \cite{Huang:2015p39241} we estimated the physical properties of the confirmed \lya\ emitter presented in Section~\ref{sec:rxj1347_1037}.
An important aspect of this study was the availability of ancillary Spitzer photometry from SURFS-UP \citep{Bradac:2014p34085}.
Photometry in the rest-frame optical falling in the Spitzer IRAC infrared bands for high-redshift galaxies, has proven to be an important part of reliably predicting the physical properties of (high-redshift) galaxies through spectral energy distribution fitting
\citep[e.g.,][]{Schaerer:2010p41944,Labbe:2013p34884,Smit:2014p37826,Smit:2014p36152,Huang:2015p39241,Finkelstein:2015p41579,Wilkins:2015p41785}
Furthermore, fixing the redshift of the spectral energy distributions when fitting to photometry, can also be used as a test of the validity of potential low-redshift contaminants. 
If for instance the best-fit low-redshift model predicts a dusty red and old stellar population, it would be very unlikely to see strong [OII] emission, therefore making a high-redshift \lya\ scenario more likely \citep{Finkelstein:2013p32467,Coe:2013p26313}. Similar arguments can be used to rule out other line-emitting low-redshift contaminants.
Fixing the redshift of high-redshift sources behind massive clusters, like the ones presented in the current study behind the GLASS clusters, is not only important for the study of individual sources and high-redshift galaxy populations.
Knowing the redshift, i.e. the luminosity distance to any object, precisely, especially if it is multiply lensed, is also very valuable for lens modeling of the foreground clusters \citep[e.g.][]{Coe:2013p26313,Coe:2014p35275,Zitrin:2014p36332}.
Lastly, the sizes of high-redshift galaxies have also been shown to provide useful information about the environment and epoch they inhabit \citep{Ono:2013p26883,CurtisLake:2014p37045,Holwerda:2015p35592}.

As exemplified by the objects described in Section~\ref{sec:individualobj}, we expect several of the line emitters presented in Table~\ref{tab:dropouts_EL} to be true \lya\ emitters. However, as shown in Section~\ref{sec:stats}, we also expect a considerable fraction of the objects to be contaminants resulting from either low-redshift line emitters or spurious line detections in the GLASS spectra. We therefore consider the current sample to be premature for a full spectral energy distribution study including comparing their inferred physical properties given the expected low purity of the presented sample.
An in-depth study of the purely photometric Gold and Silver samples is beyond the scope of this work, as it would benefit greatly from the inclusion of ancillary Spitzer photometry and a detailed knowledge of the purity and completeness functions of the complex heterogeneous photometric samples.
We defer such a study to a future publication when the full HFF data is available, and the SURFS-UP Spitzer data have been analyzed. 

\section{Stacking the \lowercase{$z\gtrsim7$} spectra}\label{sec:stack}

The \lya\ emission line has so far been the main rest-frame UV
emission line used for spectroscopic confirmation of high (and to some
extent low-redshift) galaxies due to its characteristic spatial
profile, strength, and accessibility in the optical/NIR.
However, over the last several years, searches for and studies of
other rest-frame UV lines like  CIV$\lambda$1549\AA{} 
and CIII]$\lambda$1909\AA{} have shown that these are potentially strong
enough to complement Ly$\alpha$ in the identification and studies of
galaxies close to the epoch of reionization, where the optical depth to
\lya\ is epxected to be high due to the increasingly neutral IGM
\citep[e.g.,][]{Stark:2010p27668,Stark:2011p27664,Stark:2013p28962,Stark:2014p36777,Schenker:2014p35145}.  Observations of sources at redshifts below 3
\citep[][]{Shapley:2003p36937,Erb:2010p36933,Stark:2014p36777,Rigby:2015p41811}
suggest that, indeed, CIII] (which is not affected by the neutral
hydrogen in the IGM), might be strong enough for detection with
current facilities.  As a `proof of concept' two detections of CIII]
\citep{Stark:2015p36807} and one of CIV \citep{Stark:2015p39317} in
\lya\ emitters at $z>6$ have recently been presented. 
The CIII] emission is generally faint and therefore difficult to detect, 
and current estimates of $f_{\textrm{CIII]}} / f_\textrm{\lya}$ at $z>6$ from
\cite{Stark:2014p36777} are $\lesssim0.2$.
In fact, \cite{Zitrin:2015p39322} searched for CIII] in a small sample of 
high-redshift galaxy candidates but were unable to confirm any CIII] emission.
In Table~\ref{tab:dropouts} objects from this search, overlapping with our samples
have been marked by `*'. 

Thus, even though it is challenging, a systematic search for the CIII]
and CIV lines at redshifts close to the epoch of reionization would
be extremely valuable to improve our understanding of the IGM and of
the galaxies themselves.  Detection of these UV lines would enable
photoionization modeling of line strengths, improving our
understanding of the stellar populations and metallicities of these
galaxies \citep{Stark:2014p36777} and ultimately of their output of
ionizing photons.

\begin{figure}
\begin{center}
\includegraphics[width=0.50\textwidth]{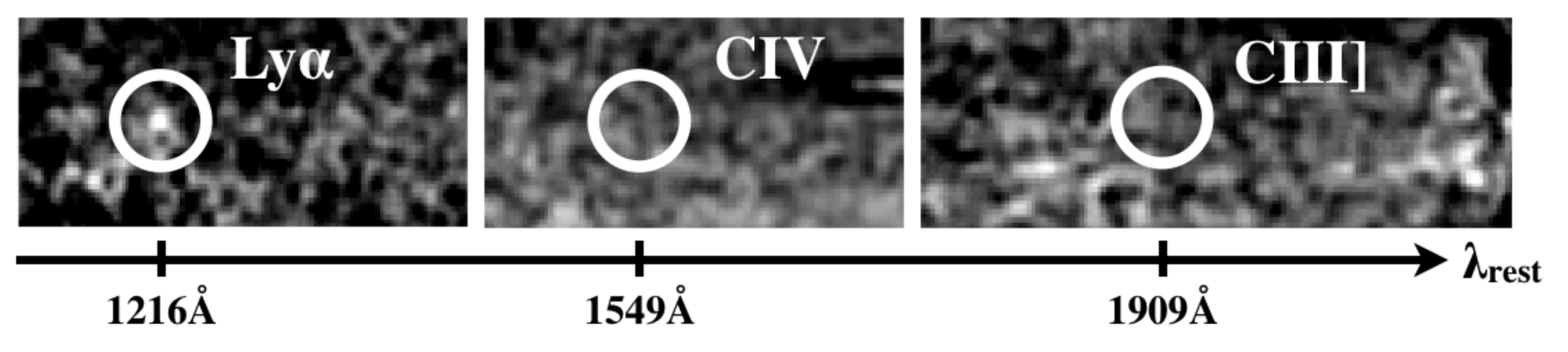}
\caption{Cutouts of the stacked two-dimensional spectrum of the Gold\_EL emission line objects listed in Table~\ref{tab:dropouts_EL}.
The mean redshift of the shown stack is $\langle z\rangle=7.2$. 
The stacked \lya\ marked by the white circle on the left, appears compact and is not resolved (see top panel of Figure~\ref{fig:profiles}).
There is no detection of a continuum break red-wards of the \lya\ line.
The stacks show no detection of CIV$\lambda$1549\AA{} and CIII]$\lambda$1909\AA{} as indicated by the two right-most white circles.}
\label{fig:ELstack}
\end{center}
\end{figure}

As shown in Figure~\ref{fig:fluxlimits} the GLASS spectra reaches
limiting line fluxes of $\sim5\times10^{-18}$~erg/s/cm$^{2}$. At
this depth, we do not detect CIII] or CIV emission in the Gold\_EL and
Silver\_EL samples.
To improve the S/N we stacked the GLASS spectra of the various samples
to search for potential CIV and CIII] emission.
Figure~\ref{fig:ELstack} shows the relevant parts of the rest-frame
stacks (assuming the \lya\ redshifts listed in
Table~\ref{tab:dropouts_EL}) of the Gold\_EL sample.  
The position of \lya\, CIV and CIII] is marked by the white circles.  
We do not detect any continuum break red-wards of the \lya\ emission
in the stacks of the dropouts presented here.
We do not detect any CIV and CIII] emission either.
From the Gold\_EL $\langle z\rangle=7.2$ stack we estimate the flux ratio limits between
the rest-frame UV emission lines to be:
\begin{eqnarray}
f_{2\sigma\textrm{lim., CIV}} / f_\textrm{\lya} 		&\lesssim& 0.32 \\
f_{2\sigma\textrm{lim., CIII]}} / f_\textrm{\lya}   		&\lesssim& 0.23    \; ,
\end{eqnarray}
where we used 2$\sigma$ limiting fluxes for CIV and CIII] estimated on
the stacked spectra. The $f_\textrm{\lya}$ was also measured directly
from the stack as described in Section~\ref{sec:flimandEW}.  These
upper limits agree well with the $f_\textrm{CIV} / f_\textrm{\lya}$
ratios presented by \citep{Stark:2015p39317} and the $f_\textrm{CIII]}
/ f_\textrm{\lya}$ ratios for low-metallicity objects at intermediate
and high $z$ presented by
\cite{Shapley:2003p36937,Erb:2010p36933,Stark:2014p36777,Stark:2015p36807}
which ranges from 0.1 to 0.6 and 0.1 to 0.3, respectively.

In addition to allowing us to look for rest-frame UV lines, the
stacked high-S/N spectra enable us to further test the hypothesis that
the majority of the emission is \lya\ at high redshift.  Following a
line of arguments similar to what was presented in
Section~\ref{sec:individualobj}, we would expect significant [OIII]
emission in the GLASS spectra, should the majority of the sources be
low redshift [OII] emitters, with the photometric dropout caused by
the 4000\AA{} break.
The stack does \emph{not} show any flux
excess at 5007\AA\ rest-frame, and the line emission is therefore
unlikely to come from low-redshift [OII] emission.

Including the Silver\_EL sample in the GLASS stack (or stacking the Silver\_EL objects
separately) does not change any of the above conclusions.

\section{The Spatial Extent of Ly$\alpha$ at $\lowercase{z}\gtrsim7$}\label{sec:spatialextent} 

A unique feature of the GLASS spectra is the high angular
resolution, owing to the sharp \emph{HST} point spread function (PSF)
and the magnification by the foreground clusters. Thus, GLASS provides
an opportunity to measure, for the first time, the angular extent of
the \lya\ emission.  Spatial information is preserved in the extracted
two-dimensional spectra and the spatial extent of emission lines can in principle
be estimated.  

\begin{figure*}
\begin{center}
\includegraphics[width=0.49\textwidth]{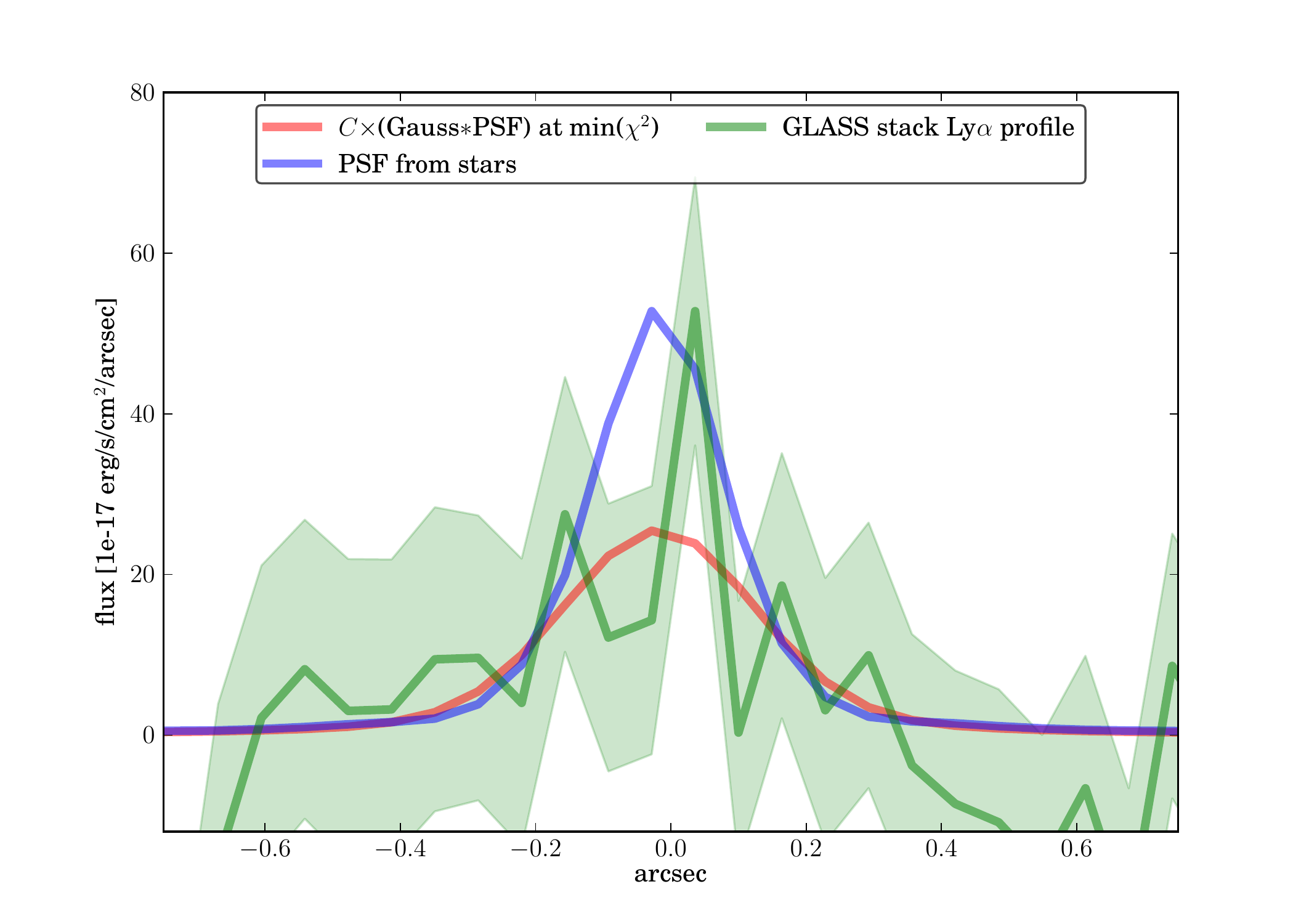}
\includegraphics[width=0.49\textwidth]{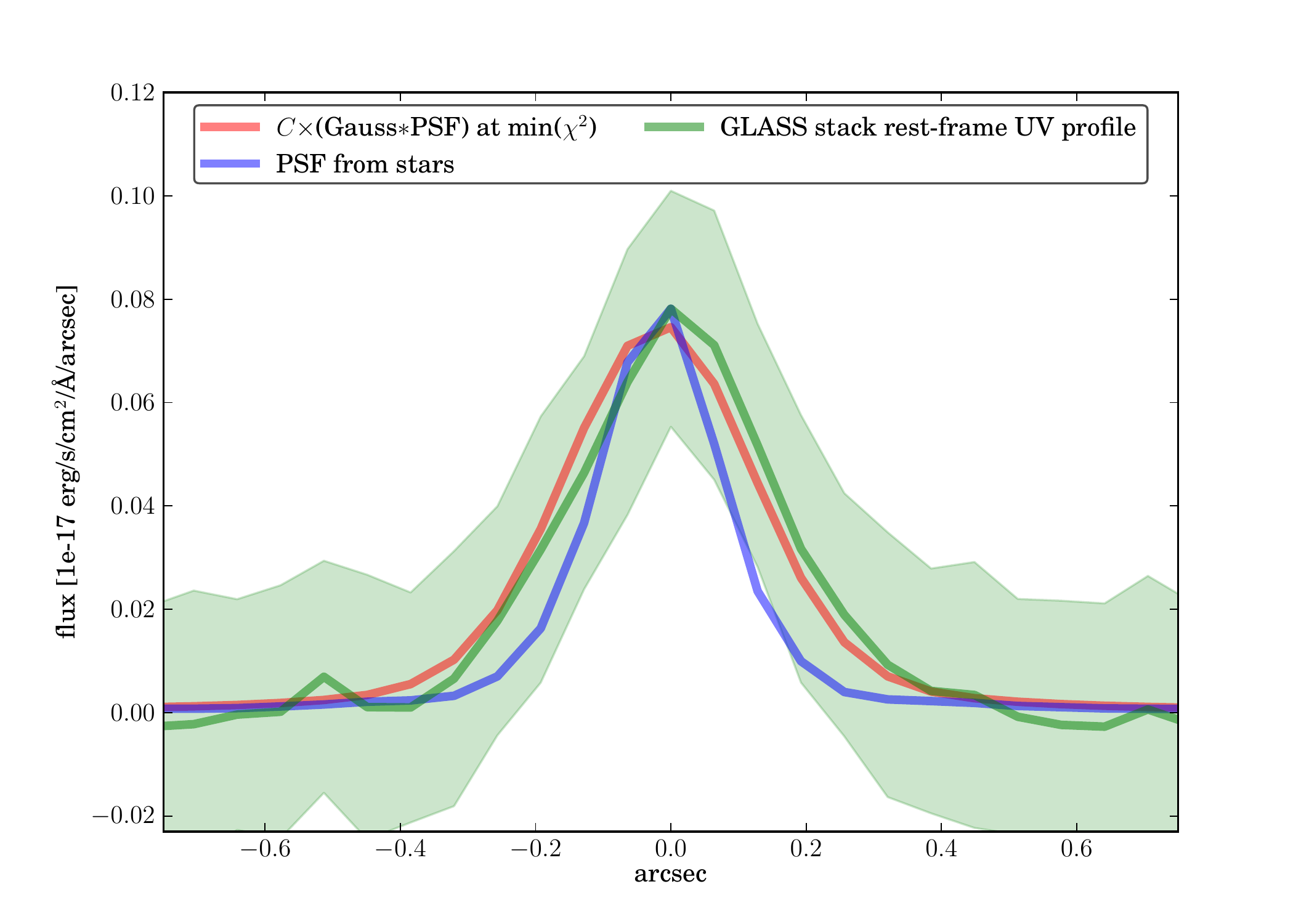}\\
\includegraphics[width=0.49\textwidth]{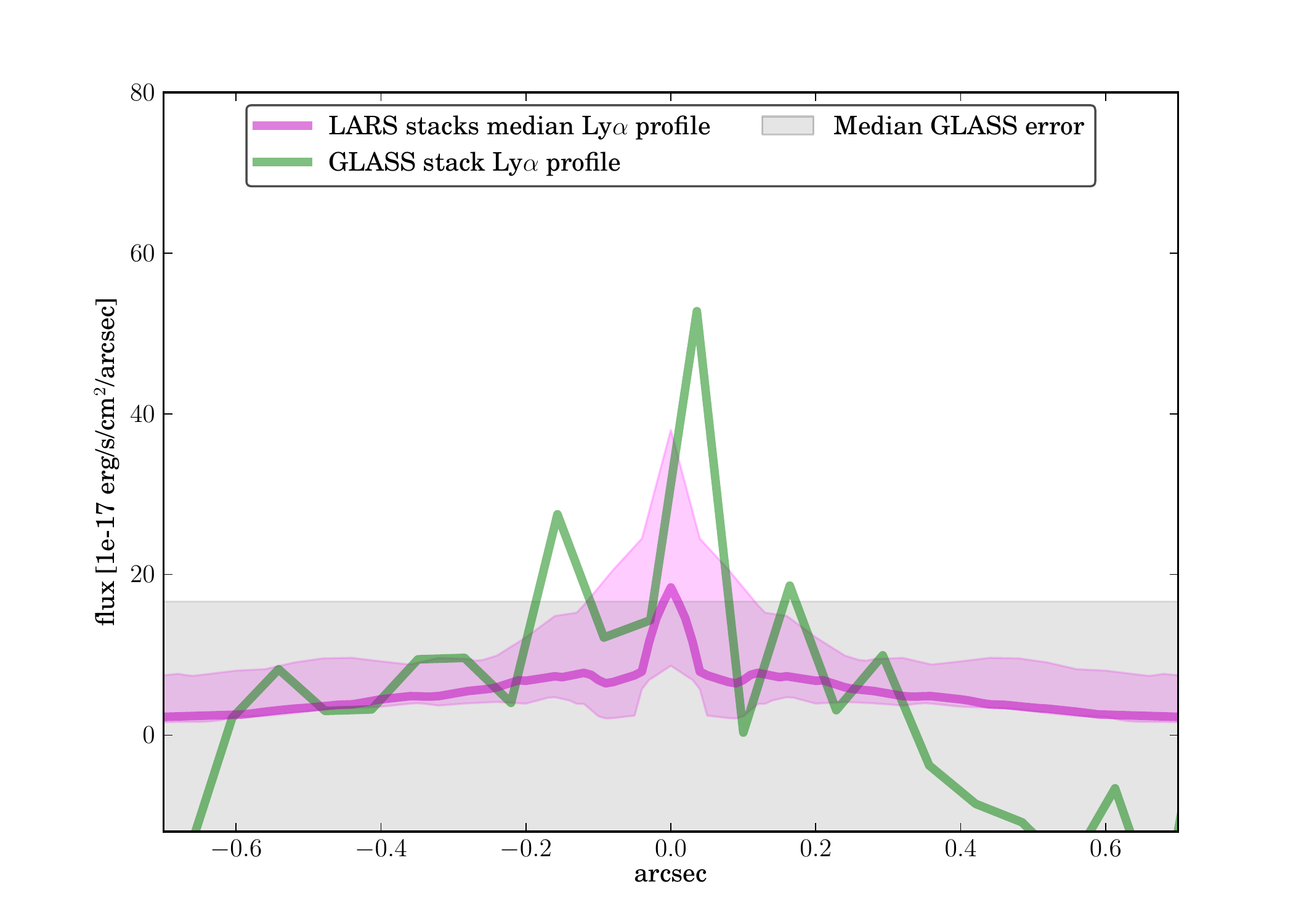}
\includegraphics[width=0.49\textwidth]{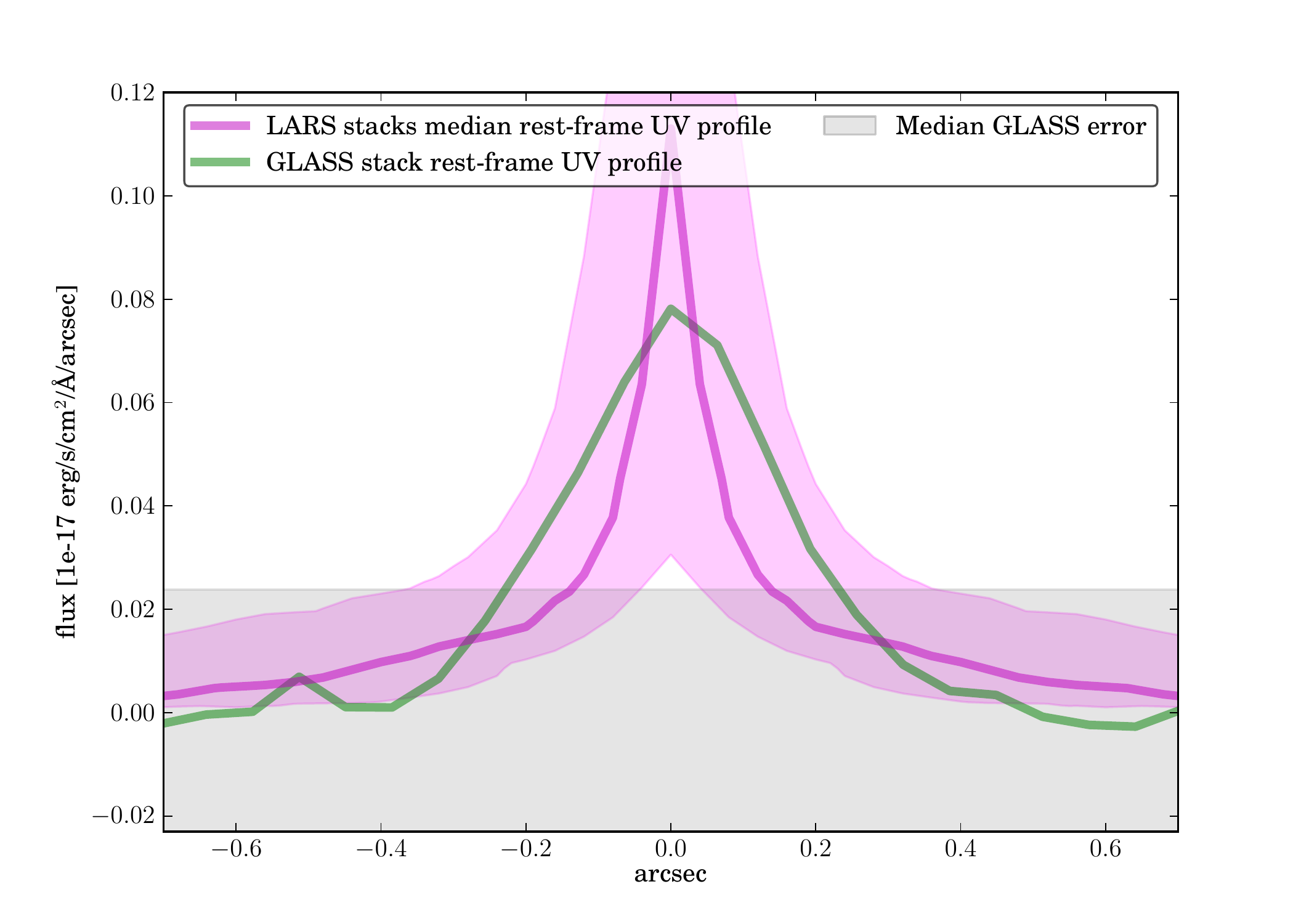}
\caption{The spatial profiles of the stacked Gold\_EL \lya\ emitters at $\langle z\rangle=7.2$ (left, green curve) and their observed NIR (rest-frame UV) direct images (right, green curve).
In the top panels these profiles are compared to the PSF represented by the spatial profile of stars (blue curves) in the GLASS filed-of-views.
The red curves show the convolved PSF (multiplied by a constant $C$) that minimizes the $\chi^2$ between the PSF and the data. 
Both the \lya\ and the rest-frame UV profiles from the GLASS stacks are unresolved.
Hence, taken at the face value, there is no evidence that the spatial extent of \lya\ is more extend than the UV light in the GLASS stack.
The bottom panels show a comparison to the median LARS \lya\ and rest-frame UV profiles at $z=7.2$ cf. Section~\ref{sec:LARS}.
The shaded area around the profiles shows the 1$\sigma$ spread of the individual profiles.
The median error on the GLASS stacks (from the top panel's green shaded region) is represented by the gray band.
Based on this comparison, we conclude that the extended \lya\ emission surface brightness typical of lower-redshift Lyman break galaxies is too faint to be detected in this GLASS stack.}
\label{fig:profiles}
\end{center}
\end{figure*}

From the $\langle z\rangle=7.2$ \lya\ stack presented in
Section~\ref{sec:stack} we extracted the spatial profile of the
\lya\ emission, by collapsing a 20\AA{} window centered on the line in
the dispersion direction (roughly the width of circle marking 
the \lya\ emission in Figure~\ref{fig:ELstack}).  
The resulting profile is shown in green in
the top left panel of Figure~\ref{fig:profiles}.  In the same figure, the
PSF of the data is shown in blue.  The PSF was obtained by extracting
the spatial profile from a stack of stellar spectra from the GLASS
observations.  We also extracted the spatial profile of the stacked
NIR (rest-frame UV) images representing the continuum light profile of
the stack. This profile is shown in the top right panel of
Figure~\ref{fig:profiles}.  To determine whether the extent of \lya\
and the rest-frame UV continuum deviates from the PSF, we modeled each
of them as a Gaussian convolution of the PSF (G$\ast$PSF) times a
constant $C$.  By sampling values of the standard deviation of the
Gaussian convolution kernels ($\sigma_\mathcal{G}$) and estimating the
minimum $\chi^2$ as
\begin{eqnarray}
\chi^2 	&=& \sum_i \frac{\left(P_{i,\textrm{\lya}} - C \times P_{i,\mathcal{G}\textrm{$\ast$PSF}}\right)^2}{\sigma^2_{i,\textrm{\lya}}} \;,\\
C		&=& \sum_i \frac{P_{i,\textrm{\lya}} \times P_{i,\mathcal{G}\textrm{$\ast$PSF}} }{\sigma^2_{i,\textrm{\lya}}} \Bigg/
			\sum_i \frac{P_{i,\mathcal{G}\textrm{$\ast$PSF}}^2}{\sigma^2_{i,\textrm{\lya}}} \;,
\end{eqnarray}
we can quantify the deviation of the spatial profiles from the PSF.
In the expressions above, $P_{i,\textrm{\lya}}$ and
$P_{i,\mathcal{G}\textrm{$\ast$PSF}}$ refer to the $i$'th pixel in the
spatial profiles of the \lya\ (or rest-frame UV) and the convolved
PSF, respectively. 
The $\sigma^2_{i,\textrm{\lya}}$ is the variance on the
\lya\ (or rest-frame UV) profile and the constant $C$ is minimized
analytically by setting $\partial \chi^2 /\partial C = 0$.  The
profiles of the convolved PSF minimizing $\chi^2$ are shown as the red
curves in Figure~\ref{fig:profiles} and correspond to Gaussian
convolution kernels with $\sigma_\mathcal{G} = 1.0^{+0.8}_{-1.0}$
pixels and $\sigma_\mathcal{G} = 1.4^{+0.7}_{-0.6}$ pixels for the
\lya\ and rest-frame UV spatial profiles, respectively.  Hence, both
profiles are only marginally resolved, and there is no indication that
the \lya\ is more extended than the UV light in the stack.  

Several authors
\citep[e.g.,][]{Steidel:2011p35184,Finkelstein:2011p36860,Matsuda:2012p36912,Momose:2014p35332,Wisotzki:2015p41489}
have shown that the Ly$\alpha$ emission is 5-15 times more
extended than the continuum emission at redshifts $z\lesssim6$.
Taking the results from GLASS at the face value, the $\langle
z\rangle=7.2$ would seem to indicate more compact emission than at
lower redshift.  However, as described in the following section, the
sensitivity of the GLASS stack is insufficient to detect the extended
\lya\ emission and we are thus most likely only seeing the high surface brightness core.

\subsection{Comparison to the LARS Sample}\label{sec:LARS}

To quantify our ability to detect extended \lya\ emission in our GLASS
$\langle z\rangle=7.2$ stack, we carry out a systematic comparison
with the rest-frame UV and \lya\ emission in the \lya\ reference
sample (LARS) galaxies presented by
\cite{Hayes:2013p39968,Hayes:2014p39965} and \cite{Ostlin:2014p39933}.

The LARS sample consists of 14 low-redshift (all with $z < 0.2$) star
forming \citep[SFR$_\textrm{FUV} > 0.5 M_\odot$/yr;][]{Hayes:2014p39965} 
Lyman break galaxy analogs. It has been observed extensively with 
HST from the UV to the optical \citep[cf. Table~4 in][]{Ostlin:2014p39933}.
The wealth of data has enabled the creation of high-resolution
rest-frame UV, \lya, H$\alpha$ and H$\beta$ maps, suitable for
comparison with low, as well as high redshift counterparts.  Following
the prescription outlined by \cite{Guaita:2015p39425}, we simulated
the observations of 12 of the LARS galaxies redshifted to $z=7.2$, the
mean redshift of the GLASS stack (we also simulated $z=7$ and $z=7.4$
to verify the redshift sensitivity and found it to be negligible).  We
did not include LARS04 and LARS06, as they do not show any
\lya\ emission within the HST field of view. 

In practice, the HST images of the LARS galaxies were resampled to a
$0\farcs04$ pixel scale (similar to the $0\farcs06$ used for the GLASS interlacing), 
corresponding to $\sim0.2$kpc at $z=7.2$,
fixing the physical size and preserving flux as outlined by \cite{Guaita:2015p39425}.  
After resampling, the \lya\ and rest-frame UV emission was isolated via continuum subtraction using the `continuum throughput normalization' (CTN) factor as described by \cite{Hayes:2005p40043,Hayes:2009p39977}.
The continuum-subtracted emission maps were then scaled based on luminosity distance and surface brightness dimming,
\citep[e.g.,][]{Bouwens:2004p31890}.  
For more details on the LARS sample and the high redshift simulations
we refer the reader to \cite{Hayes:2013p39968,Hayes:2014p39965,Ostlin:2014p39933} and
\cite{Guaita:2015p39425}.

The individual spatial profiles of the LARS galaxies were produced by
collapsing the UV and \lya\ images after an arbitrary rotation around
the rest-frame UV emission centroid.  For each galaxy, these
individual profiles were aligned and combined to a main spatial UV and
\lya\ profile for each of the 12 galaxies.  This essentially
corresponds to stacking 12 samples of $N$ `different' galaxies.  The
medians of these 12 rest-frame UV and \lya\ profiles are shown at a
simulated redshift of 7.2 in magenta in the bottom panels of
Figure~\ref{fig:profiles}.  The shaded regions show the 68\% spread of
the 12 profiles.  Comparing these profiles to the GLASS rest-frame UV
and \lya\ taking the median error on the GLASS stacks into account
(horizontal gray band), it is clear that in most cases the extended
\lya\ emission would not be detectable in the GLASS stack. Only the central 
high surface brightness peak appears detectable with the GLASS
sensitivity. In this simulation we have neglected the effect of
magnification. 
If we scale up the LARS galaxies by the 
average linear magnification $\langle \sqrt{\mu}\rangle \sim 1.5\pm1.0$
the effect is unchanged.
If we only consider the LARS LAEs (EW$_\textrm{\lya}>20$\AA),
the conclusions do not change either.

A similar conclusion is reached by comparing the surface brightness
profiles and uncertainties of the GLASS stack with the \lya\ surface
brightness profiles presented by \cite{Momose:2014p35332} and \cite{Wisotzki:2015p41489},
even though both of these studies are only considering redshifts lower than 7.

\section{Summary}
\label{sec:conc}

In this paper we have presented a systematic search for \lya\ emission
from galaxies at the epoch of reionization. 
We have analyzed the GLASS spectroscopy of 159 photometrically pre-selected
$z\gtrsim7$ galaxy candidates lensed by the first six clusters
observed as part of GLASS. 
Our main results can be summarized as follows:

\begin{enumerate}
\item  From visual inspection of all 159 spectra, we find emission features
consistent with being \lya\ in 24 objects. Assuming the lines are all \lya,
the mean redshift of the emission line sample is $7.3$ ($7.2$ and $7.4$ for the Gold\_EL 
and Silver\_EL sample, respectively), with a few candidates
above $z\sim8$. By comparison with automatic line detection results,
we estimate the completeness of this sample to be 
40-100\% with a purity of 60-90\%. Deeper spectroscopic follow-up is
needed to improve the estimates of completeness and purity.

\item One of the candidates has been confirmed spectroscopically with 
DEIMOS on Keck. The long wavelength coverage of the GLASS grism allows
us to confirm that the line is indeed \lya\ and not [OII] at lower
redshift.

\item The most compelling candidate at $z>8$, 
is detected independently in both of the G102 grism spectra 
(and marginally in one of the G141 spectra). The total
line flux is $1.0\pm0.3 \times10^{-17}$erg/s/cm$^{2}$ and the
wavelength of the emission line corresponds to \lya\ at $z=8.10$, consistent with the
photometric redshift. Follow-up spectroscopy or deeper imaging is
needed to confirm the candidate.

\item The number of emission line detections is consistent with the expectations based 
on our knowledge of the \lya\ emission probability for Lyman Break
Galaxies at $z\sim7$, although the uncertainties are large. The full
analysis of the GLASS sample together with a homogenous selection
based on the Hubble Frontier Field HST (and SURFS-UP Spitzer IRAC)
imaging dataset is necessary to
carry out a more quantitative analysis, and measure the \lya\ optical
depth to $z\sim7$ and $z\sim8$ sources.

\item From a stack of the most promising \lya\ emitters we derive the 
spatial profile of \lya\ at $\langle z\rangle=7.2$. The stacked \lya\
profile is comparable in size to that of the continuum UV emission,
and only marginally resolved with respect to the point spread
function. Diffuse \lya, if present, is below our surface brightness
detection limit. We show that this is consistent with the properties
of galaxies at lower redshift, by simulating observations of the low
redshift LARS Lyman break galaxy analogs at $z=7.2$.

\item We do not detect any CIV or CIII] emission 
in the $\langle z\rangle=7.2$ stack, down to a 1$\sigma$ \lya\ limit of 
$2\times10^{-18}$erg/s/cm$^{2}$ (not corrected for magnification).

\end{enumerate}

In conclusion, our search has confirmed that \lya\ is getting harder
and harder to detect as we approach the epoch when the universe is in
large part neutral. The space-based data guarantee that this is not
due to the unfortunate coincidence of sky emission lines and redshifted
\lya.  However, the extreme faintness of these lines requires even deeper
follow-up spectroscopy, in order to make progress, by improving our
estimates of completeness and purity. Higher spectral resolution data
would also help remove contamination by detecting the
characteristically asymmetric \lya\ profile. We have published this
first sample as quickly as possible with the aim of fostering
follow-up efforts.

\section{Acknowledgments}

This paper is based on observations made with the NASA/ESA Hubble Space Telescope, obtained at STScI.
Support for GLASS (HST-GO-13459) was provided by NASA through a grant from the Space Telescope Science Institute, which is operated by the Association of Universities for Research in Astronomy, Inc., under NASA contract NAS 5-26555. 
We are very grateful to the staff of the Space Telescope for their assistance in planning, scheduling and executing the observations. 
BV acknowledges the support from the World Premier International Research Center Initiative (WPI), MEXT, Japan and  the Kakenhi Grant-in-Aid for Young Scientists (B)(26870140) from the Japan Society for the Promotion of Science (JSPS).
AH acknowledges support by NASA Headquarters under the NASA Earth and Space Science Fellowship Program - Grant ASTRO14F-0007. MB, AH, and KH also acknowledge support by NASA through an award issued by
JPL/Caltech and from STScI via HST-AR-13235, HST-GO-13177, and special funding as part of the HST Frontier Fields program conducted by STScI.
TAJ acknowledges support from the Southern California Center for Galaxy Evolution through a CGE Fellowship.
This work utilizes gravitational lensing models produced by PIs Brada\v{c}, Ebeling, Merten \& Zitrin, Sharon, and Williams funded as part of the \emph{HST} Frontier Fields program conducted by STScI. STScI is operated by the Association of Universities for Research in Astronomy, Inc. under NASA contract NAS 5-26555. The lens models were obtained from the Mikulski Archive for Space Telescopes (MAST).

\begin{appendix}
\section{GLASS Confirmation of \lowercase{$z=6.1$} and \lowercase{$z=6.4$} \lya\ emitters in \Rtwentytwo\ and \Mseven}
\label{sec:VandBB}

Currently, two multiple imaged systems at high redshift have been spectroscopically confirmed in the six clusters analyzed in this work. In \Rtwentytwo\ \cite{Boone:2013p35081} and \cite{Balestra:2013p35083} presented spectroscopy of a  multiple imaged system at $z=6.1$ and \cite{Vanzella:2014p33637} confirmed a system in \Mseven\ at $z=6.4$.
In Table~\ref{tab:BBV} we list the individual components of these systems detected in the GLASS data. 
Due to their $z\sim6$ redshifts, these objects were not included in the sample of dropouts studied in this paper.
As expected, the main components have photometric redshift estimates around 6 as shown by the 1$\sigma$ intervals from EA$z$Y quoted in Table~\ref{tab:BBV}.

\begin{turnpage}

\tabletypesize{\tiny} \tabcolsep=0.11cm
\begin{deluxetable*}{ccccccccccccc} \tablecolumns{13}
\tablewidth{0pt}
\tablecaption{The Multiple Imaged Sources from \cite{Boone:2013p35081,Balestra:2013p35083} and \cite{Vanzella:2014p33637}}
\tablehead{
 \colhead{Cluster} & \colhead{ID} & \colhead{ID} & \colhead{R.A.} & \colhead{Dec.} & \colhead{F105W$_\textrm{CLASH}$} & \colhead{$z_\textrm{EA$z$Y}$} & \colhead{P.A.} & \colhead{$\lambda_\textrm{\lya}$} & \colhead{$z_\textrm{\lya}$} & \colhead{EW$_\textrm{\lya}$}  & \colhead{$f_\textrm{\lya}$} & \colhead{$\mu$} \\
 \colhead{} & \colhead{GLASS} & \colhead{CLASH} & \colhead{[deg]} & \colhead{[deg]} & \colhead{[ABmag]} & \colhead{$1\sigma$ range} & \colhead{[deg]} & \colhead{[\AA]} & \colhead{} & \colhead{[\AA]}  & \colhead{[1e-17 erg/s/cm$^2$]} & \colhead{}
}
\startdata 
\Rtwentytwo\  &	00699 & 01291 & 342.19089  &   -44.53746  &    $24.82 \pm 0.04$  & [5.807,5.940]    & 053, 133 & 8641, 8613  &  6.106,  6.083   &   $ 45 \pm 8 $, $ 46 \pm 8 $      			&  $ 5.54 \pm 1.01 $, $ 5.59 \pm 0.94 $   				&     $4.26\pm2.85$   	\\
\Rtwentytwo\  &	00845 & 01154 & 342.18104  &   -44.53463  &    $25.09 \pm 0.06$  & [0.065,0.183]    & 053, 133 & 8638, 8642  &  6.104,  6.107   &   $ 53 \pm 11 $, $ 63 \pm 14 $  			&  $ 5.03 \pm 1.01 $, $ 5.99 \pm 1.29 $  				&     $3.74\pm5.32$ 	\\
\Rtwentytwo\  &	01131 & 00847 & 342.18904  &   -44.53002  &    $24.29 \pm 0.03$  & [0.912,1.053]    & 053, 133 & 8635, 8640  &  6.101,  6.105   &   $ 68 \pm 4^\star $, $ 22 \pm 4^\star $      	&  $ 13.63 \pm 0.74^\star $, $ 4.44 \pm 0.83^\star $   	&     $4.16\pm8.35$   	\\
\Rtwentytwo\  &	01752 & 00401 & 342.17130  &   -44.51981  &    $25.94 \pm 0.08$  & [5.789,5.994]    & 053, 133 & 8650, 8633  &  6.113,  6.100   &   $ 77 \pm 24 $, $ 46 \pm 20 $  			&   $ 3.35 \pm 1.02 $, $ 1.99 \pm 0.88 $   				&     $2.03\pm1.2$     	\\
\Mseven\        &	00846 & 01730 & 109.40773  &     37.74274  &   $26.42 \pm 0.11$  & [5.745,5.989]    & 020, 280 & 8985, 8985  &  6.389,  6.389   &   $ 348 \pm 57^\star $, $ 25 \pm 24^\star $	&   $ 10.00 \pm 0.69^\star $,$ 0.72 \pm 0.67^\star $   	&   $10.96\pm13.39$  	\\
\Mseven\  	  &	02068 & 00859 & 109.40907  &     37.75469  &   $26.34 \pm 0.16$  & [5.846,6.084]    & 020, 280 & 8980, 8987  &  6.385,  6.391   &   $ 78 \pm 33 $, $ 134 \pm 45 $			&   $ 2.11 \pm 0.86 $, $ 3.62 \pm 1.16 $   				&    $8.15\pm6.63$      	
\enddata
\tablecomments{$^\star$Flux and EW estimates are tentative due to high contamination in the GLASS spectra.
$\mu$ gives the magnifications of the HFF clusters obtained as described in Section~\ref{sec:flimandEW}.
}
\label{tab:BBV}
\end{deluxetable*}


\end{turnpage}

\begin{figure}
\begin{center}
\includegraphics[width=0.49\textwidth]{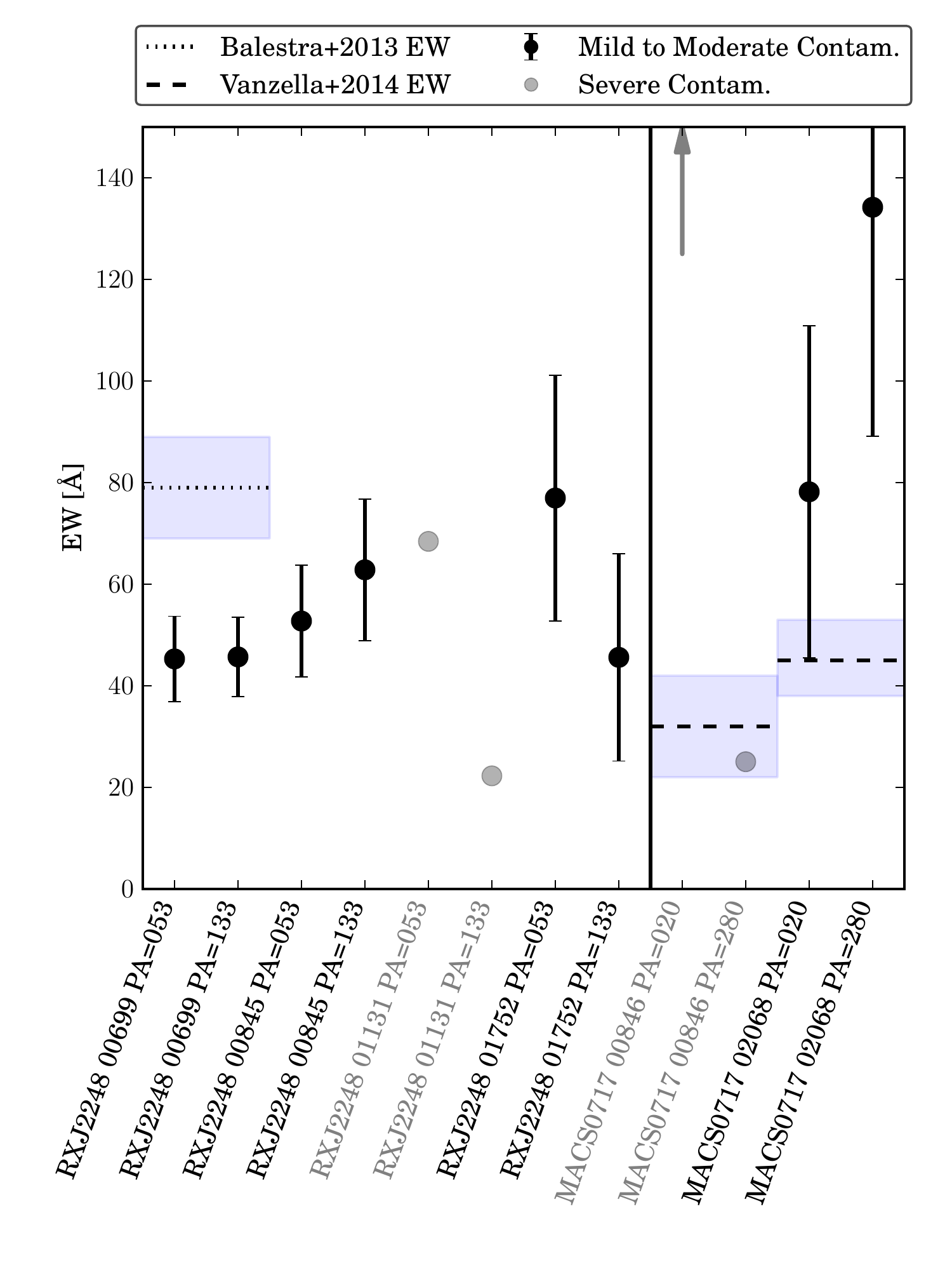}
\caption{The equivalent widths estimated from the detections of \lya\ in the individual GLASS spectra of the four main images of the multiple imaged system at $z=6.1$ from \cite{Boone:2013p35081} and \cite{Balestra:2013p35083} (left part) and the two images of the $z=6.4$ \lya\ emitter from \cite{Vanzella:2014p33637} (right part) listed in Table~\ref{tab:BBV}.
The equivalent widths estimated by \cite{Balestra:2013p35083} and \cite{Vanzella:2014p33637} are marked by the dotted and dashed lines, respectively. The shaded regions indicate the quoted uncertainties.
The gray dots represent GLASS spectra with severe contamination at the location of \lya\ making the GLASS equivalent width estimates only tentative.
The arrow indicates that the tentative estimate of MACS0717\_00846 at PA=020 is outside the plotted range.
The equivalent width estimates from the GLASS spectra generally agree with each other.
There is marginal agreement with the \cite{Vanzella:2014p33637} estimates, whereas the GLASS equivalent widths are generally lower than the estimate from \cite{Balestra:2013p35083}.}
\label{fig:BBVew}
\end{center}
\end{figure}

We list the apparent position of the \lya\ emission line in the GLASS G102 spectra, and the corresponding redshift. In a few cases,  significant flux contamination at the location of \lya\ make the flux estimates, and hence the equivalent widths, only tentative.
In Figure~\ref{fig:BBVew} we compare the equivalent widths from GLASS with the values quoted in the literature.

Combining the equivalent width for the pairs of GLASS spectra we find EW$_\textrm{\lya}=45\pm6$\AA, EW$_\textrm{\lya}=58\pm9$\AA\ and EW$_\textrm{\lya}=61\pm16$\AA\ for RXJ2248\_00699,  RXJ2248\_00845, and  RXJ2248\_01752, respectively. These all agree within 1$\sigma$ of each other. 
However, when comparing RXJ2248\_00699 GLASS equivalent widths with the equivalent width published by \cite{Balestra:2013p35083}, 
we find a discrepancy of 5.7$\sigma$. If we combine the 6 (3 sources x 2 PAs) 
GLASS equivalent widths we get EW$_\textrm{\lya}=55\pm5$ and this discrepancy shrinks to 4.8$\sigma$.
For MACS0717\_02068 we find a combined \lya\ equivalent width of EW$_\textrm{\lya}=106\pm26$\AA{} which is within 2.4$\sigma$ of the 45\AA{} estimated for this source by \cite{Vanzella:2014p33637}.

In summary, the individual GLASS estimates are self-consistent but seem to disagree somewhat with the ground-based measurements for the \Rtwentytwo\ system.
This discrepancy might also be explained by a combination of slit losses and the challenging modeling of the contamination and background subtraction in the GLASS spectra.

\section{The Automatic Line Detection Significance for the Emission Line Samples}
\label{sec:plinevals}

In Table~\ref{tab:plinevals} we quote the line detection significance
from the Bayesian line detection algorithm described in Section~\ref{sec:pline} and \cite{Maseda:2015p39932}
applied to the emission
line sample from Table~\ref{tab:dropouts_EL}. 
The quoted values correspond to the maximum
$p$-values in the $\pm$ 50\AA{} range around the
visually detected emission lines given in the 
`$\lambda_\textrm{lines} \pm50$\AA' column
of Table~\ref{tab:plinevals} (and Table~\ref{tab:dropouts_EL}).


\tabletypesize{\scriptsize} \tabcolsep=0.2cm
\begin{deluxetable*}{cccccclll} \tablecolumns{9}
\tablewidth{0pt}
\tablecaption{Automatic Line Detection Significance of the Emission Line Samples from Table~\ref{tab:dropouts_EL}}
\tablehead{
 \colhead{Cluster} & \colhead{ID} & \colhead{R.A.} & \colhead{Dec.} & \colhead{P.A.} & \colhead{$\lambda_\textrm{lines}$} & \colhead{$p_\textrm{max}$} & \colhead{$\sigma$} & \colhead{  {$\sigma_\textrm{tot}^\diamond$}}  \\
 \colhead{} & \colhead{GLASS} & \colhead{[deg]} & \colhead{[deg]} & \colhead{[deg]} & \colhead{$\pm$50\AA} & \colhead{} & \colhead{} & \colhead{} 
}
\startdata 
 A2744 		& 00463 	& 3.604573038 	& -30.409357092 	& 135, 233    &   9395, \nodata 		  &       0.9976003, 0.8201783      	&        3.04,  1.34      &        3.32    \\
 A2744 		& 00844 	& 3.570068923 	& -30.403715689 	& 135, 233    &   8929, \nodata 		  &       0.9709888, 0.8801960      	&        2.18,  1.56      &        2.68     \\
 MACS1423 	& 00648 	& 215.945534620 	& 24.072435174 	& 008, 088    &   9585, \nodata 		  &       0.9899920, 0.5552243      	&        2.58,  0.76      &        2.69     \\
 MACS1423 	& 01102 	& 215.935869430 	& 24.078415134 	& 008, 088    &   9681, \nodata 		  &	      0.8870970, 0.9054205      	&        1.59,  1.67      &        2.31      \\
 MACS2129 	& 00677 	& 322.353239440 	& -7.697441500 	& 050, 328    &   9582, 9582 		  &       0.9995228, 0.9999788      	&        3.49,  4.25      &        5.50      \\
 MACS2129 	& 00899 	& 322.343220360 	& -7.693382243 	& 050, 328    &   11059, 11069 		  &       0.5700795,$\quad$\nodata 	&        0.79,\nodata &        0.79  \\
 MACS2129 	& 01516 	& 322.353942530 	& -7.681646419 	& 050, 328    &   9593, \nodata 		  &       0.9810597, 0.9870194      	&        2.35,  2.48      &        3.42        \\
 RXJ2248 		& 00207 	& 342.185601570 	& -44.547224418 	& 053, 133    &   11609,\nodata 	  &     0.9801501,$\quad$\nodata	&        2.33,\nodata &        2.33     	\\
\hline
 A2744 		& 00233 	& 3.572513845 	& -30.413266331 	& 135, 233  	& 11156,\nodata 	&	 0.7343585,$\quad$\nodata     	&        1.11,\nodata  &            1.11     \\  
 A2744		& 01610 	& 3.591507273 	& -30.392303082 	& 135, 233 	& \nodata, 8406 	&       0.9998119, 0.9999839      	&        3.73,  4.31      &              5.70      \\ 
 A2744 		& 02273 	& 3.586488763 	& -30.381334667 	& 135, 233  	& 8717, \nodata 	&       0.9890780, 0.9889105      	&        2.55,  2.54      &              3.60      \\
 MACS0717 	& 00370 	& 109.377007840 	& 37.736462661 	& 020, 280 	& 9138, \nodata 	&       0.9377069,$\quad$\nodata  	&        1.86,\nodata &               1.86      \\
 MACS1423 	& 00435 	& 215.942403590 	& 24.069659639 	& 008, 088  	& 10500,\nodata 	&       0,$\qquad\qquad$0.0000029    &  	  0, $\quad$  0  &             0            \\
 MACS1423 	& 00539 	& 215.932958480 	& 24.070875663 	& 008, 088 	& 8666, \nodata 	&       0.7680499, 0.9973003     	 	&        1.20,  3.00      &              3.23      \\
 MACS1423 	& 01018 	& 215.958132710 	& 24.077013896 	& 008, 088 	& 13702,\nodata 	&       0.9201682, 0.2224523      	&        1.75,  0.28      &              1.77      \\
 MACS1423 	& 01169 	& 215.942112130 	& 24.079404012 	& 008, 088 	& 9721, \nodata 	&       0.9998913, 0.8989955      	&        3.87,  1.64      &              4.20     \\
 MACS1423 	& 01412 	& 215.947908420 	& 24.082450925 	& 008, 088 	& 9448, \nodata 	&       0.7569422, 0.9116211       	&        1.17,  1.70      &             2.06      \\
 MACS1423 	& 01619 	& 215.935606220 	& 24.086476168 	& 008, 088 	& 9932, \nodata 	&       0.9972566, 0.9264660      	&        3.00,  1.79      &             3.49       \\
 MACS2129 	& 01182 	& 322.344533970 	& -7.688477035 	& 050, 328 	& 12145,\nodata 	&       0.9363897, 0.6058985      	&        1.85,  0.85      &              2.04      \\
 RXJ1347 		& 00627 	& 206.893075800 	& -11.760237310 	& 203, 283 	& 10750,\nodata 	&       0.9998887, 0.9416673      	&        3.86,  1.89      &              4.30      \\
 RXJ1347 		& 00997 	& 206.895685760 	& -11.754637616 	& 203, 283 	& 9467, 9463 		&       0.8859983, 0.9065139      	&        1.58,  1.68      &              2.31      \\
 RXJ1347 		& 01241 	& 206.899894840 	& -11.751082858 	& 203, 283 	& 9902, \nodata 	&       0.8757706, 0.9886466      	&        1.54,  2.53      &              2.96      \\
 RXJ2248 		& 00404 	& 342.201879400 	& -44.542663866 	& 053, 133 	& 13239,\nodata 	&       0.0000187, 0.5989891      	&        0, $\quad$ 0.84&            0.84      \\
 RXJ2248 		& 01953 	& 342.192399500 	& -44.515663484 	& 053, 133 	& \nodata, 9118 	&       0.9997395, 0.9626378      	&        3.65,  2.08      &              4.20      
\enddata
\tablecomments{The individual $p_\textrm{max}$ values represent the maximum $p$-value in the range $\lambda_\textrm{lines} \pm 50$\AA.
$^\diamond$ The $\sigma_\textrm{tot}$ is the individual significance estimates ($\sigma$) summed in quadrature.
}
\label{tab:plinevals}
\end{deluxetable*}


\section{Estimating purity and completeness}
\label{app:purcomp}

In this appendix we describe in detail the procedure adopted to
estimate the purity and completeness of the visual and automated line
detection.  
Here the completeness is defined as the number of detected lines divided by the number of lines from the selected galaxies above some flux limit.
The purity is the number of actual lines divided by the number of detected lines.
Hence, the completeness here only refers to our ability to detect emission lines among the selected galaxies, and is therefore independent of the purity and completeness of the photometric pre-selections described in Section~\ref{sec:presel}.

Let us write the notation for a single sample, using
numerical examples from the Gold sample, when necessary, and for
3$\sigma$ detections. Assuming that there are $N_w$ lines above
the flux limit that are well described in their morphology by the UV
continuum and $N_n$ that are not, the number of detections by the
automated procedure will be
\BE
N_{d,a} = c_a \cdot N_w + f_{s,a} N_G   \; ,
\EE
where $c_a$ is the completeness of the automated procedure, $f_{s,a}$
is the rate of spurious detections per spectrum, and $N_G$ is the
total number of spectra to be analyzed. The equivalent for the visual
inspection procedure will be 
\BE
N_{d,v} = c_v \cdot (N_w + N_n) + f_{s,v} N_G    \; .
\EE
For the Gold sample the numerical values are $N_G=48$, and
$N_{d,v}=8$. Analysis of the output of the automated detection
software, yields $N_{d,a}=11$ above 3$\sigma$, after removing
obviously spurious features at low grism sensitivity regions and from
contamination subtraction residual. Roughly 60\% of these lines were
not picked up by the visual classification described in
Section~\ref{sec:specsamp}.  One of these lines is the confirmed \lya\
emitter RXJ1347\_01037 (see Section~\ref{sec:individualobj}).

We know from our analysis of the spectra in empty parts of the sky
that $f_{s,a}=4/26=0.15^{+0.11}_{-0.07}$. Considering 3$\sigma$
detections we can assume that $c_a=1$ for all practical
purposes. Thus,
\BE
N_w=N_{d,a}-f_{s,a} N_G = 3.6\pm4.5   \; .
\EE
Since $N_w$ is positive, we conclude it is in the range 0 to 8. In
order to estimate the other quantities, we make use of the fact that
approximately 40\% of the lines detected automatically are also
detected visually. Thus
\BE
c_v \cdot N_w + f_{s,v} N_G = 0.4 (N_w + f_{s,a} N_G)    \; ,
\EE
which can be written as
\BE
c_v = 0.4 + 0.4*f_{s,a} N_G/N_w - f_{s,v} N_G/N_w   \; .
\EE
To solve this, we can assume conservatively that the human eye rejects
with equal probability true and false positives, so that $c_v=0.4$ and
$f_{s,v}=0.4f_{s,a}=0.06$. Alternatively we could optimistically
assume that human eye is better at removing false positives, and
therefore $f_{s,v}$ is as small as possible (0.02) and $c_v=1$. This
gives us a range of solutions for $c_v$ and $f_s$, with the visual
completeness being between 40\% and 100\%. The number of lines in the
sample not well-described by the UV flux, that could potentially be
detected is thus
\BE
N_n = \frac{N_{d,v}-f_{s,v}N_G}{c_v} - N_w   \; ,
\EE
ranging from 9 to 3 for the values of completeness estimated above. We
can use the numbers derived here to assess the total purity of the
sample, i.e. the fraction of true positives amongst the visual
detections as
\BE
1-\frac{f_{s,v} N_G}{N_{d,v}}   \; ,
\EE
which ranges between 60-90\% for the assumptions about completeness
given above.

Repeating the arguments for the Silver sample with $N_{d,v}=16$,
$N_G=87$, gives purity in the range 65-90\% for the same assumptions
about completeness.

\end{appendix}

\bibliographystyle{apj}
\bibliography{bibtexlibrary}

\end{document}